\documentclass[a4paper,twocolumn,11pt,accepted=2024-12-15]{quantumarticle}
\pdfoutput=1
\usepackage[utf8]{inputenc}
\usepackage[english]{babel}
\usepackage[T1]{fontenc}
\usepackage{amsmath}
\usepackage{hyperref}

\usepackage{tikz}
\usepackage{lipsum}

\usepackage{braket}

\usepackage{booktabs}

\begin{document}

\title{Theory of Multimode Squeezed Light Generation in Lossy Media }

\author{Denis A. Kopylov}
\email{denis.kopylov@uni-paderborn.de}
\affiliation{Department of Physics, Paderborn University, Warburger Straße 100, D-33098 Paderborn, Germany}
\affiliation{Institute for Photonic Quantum Systems (PhoQS), Paderborn University, Warburger Straße 100, D-33098 Paderborn, Germany}
\author{Torsten Meier}
\affiliation{Department of Physics, Paderborn University, Warburger Straße 100, D-33098 Paderborn, Germany}
\affiliation{Institute for Photonic Quantum Systems (PhoQS), Paderborn University, Warburger Straße 100, D-33098 Paderborn, Germany}
\author{Polina R. Sharapova}
\affiliation{Department of Physics, Paderborn University, Warburger Straße 100, D-33098 Paderborn, Germany}

\maketitle
\begin{abstract}
A unified theoretical approach to describe the properties of multimode squeezed light generated in a lossy medium is presented. 
This approach is valid for Markovian environments and includes both a model of discrete losses based on the beamsplitter approach and a generalized continuous loss model based on the spatial Langevin equation.
For an important class of Gaussian states, we derive master equations for the second-order correlation functions and illustrate their solution for both frequency-independent and frequency-dependent losses.
Studying the mode structure, we demonstrate that in a lossy environment no broadband basis without quadrature correlations between the different broadband modes exists.
Therefore, various techniques and strategies to introduce broadband modes can be considered.
We show that the Mercer expansion and the Williamson-Euler decomposition do not provide modes in which the maximal squeezing contained in the system can be measured.
In turn, we find a new broadband basis that maximizes squeezing in the lossy system and present an algorithm to construct it.

\end{abstract}

\section{Introduction}

Currently, squeezed light is of great interest in various fields of science due to its non-classical properties and simple generation. 
One of the common ways to generate squeezed light constitutes the use of single-pass optical parametric amplifiers based on parametric down-conversion (PDC) or four-wave-mixing (FWM) processes~\cite{klyshko1988book,Vogel_Welsch_book}.
Such sources provide not only efficient ways to generate biphoton pairs and squeezed states, but also can be used in more advanced circuits to generate non-Gaussian states of light~\cite{Walschaers_2021}. 

%
To efficiently generate squeezed states of light in single-pass schemes, pulsed lasers are required.
However, unlike monochromatic pumping, where the spectral components of PDC light are pairwise correlated~\cite{klyshko1988book,Caves_1982,Huttner_1990,Kolobov_1999}, pulsed pumping results in the simultaneous coupling of various monochromatic components. 
As a result, the input-output relations for the annihilation and creation operators have a form of the Bogoliubov transformation, where each output plane-wave operator is connected with a set of input operators. 
With the use of the Bloch-Messiah reduction (BMr)~\cite{Braunstein_2005,Wasilewski_2006}, the PDC input-output relations is diagonalized by introducing a new basis of broadband modes, the so-called Schmidt modes~\cite{Fabre_2020,Raymer_2020}. 
As a consequence, the output state can be interpreted as a set of independent squeezers.
A framework of independent broadband squeezers significantly simplifies the use of squeezed light in various applications, such as quantum information protocols~\cite{Brecht_2015}, quantum state engineering~\cite{Kouadou_Treps_2023, Serino_2023}, etc.  
However, for these purposes, all local oscillators, projections, or demultiplexers must be prepared in the Schmidt-mode basis, which makes finding the correct broadband basis an important task.

In order to find the Schmidt-mode basis, various approximations, such as narrowband pumping~\cite{Dayan_2007_theory} or spatially-averaged Heisenberg equations~\cite{Sharapova_2015,Pe_ina_2015,Sharapova_2018,Barral_2020}, can be used. 
These approximations allow for analytical treatment of the problem, but provide only qualitative solutions for the Bogoliubov transformation.
In turn, quantitative solutions, which take into account the spatial-ordering effects, can only be obtained numerically~\cite{Christ_2013,Horoshko_2019,Sharapova_2020,Quesada_2020}.
The main disadvantage of the above approaches is the absence of internal losses which, however, are present in any dielectric medium, e.g., due to absorption. 
Therefore, the correct quantum field dynamics should be described by the spatial Langevin equation~\cite{klyshko1988book,Vogel_Welsch_book,Caves_1987,Huttner_1992,Gruner_1996}.
When PDC is generated in transparent bulk nonlinear crystals, the absorption is small enough to be neglected, however, internal losses can be significant for structured media like waveguides, where the guided light can be lost due to scattering from surface roughness~\cite{Melati_2014}.
In addition, losses may appear due to the presence of polaritons~\cite{Klyshko_1970_polariton}, atomic transitions~\cite{Swaim_2017} or additional nonlinear processes~\cite{Rasputnyi_2022}.
Most of the existing models of squeezed light generation in a lossy medium are developed either for the bi-photon regime~\cite{Antonosyan_2014,Helt_2015, Gr_fe_2017,Banic_2022} or using the monochromatic pump approximation~\cite{klyshko1988book,Caves_1982,Caves_1987,Kumar_1984,Kolchin_2007,Raymond_Ooi_2007,Shwartz_2012,Ooi_2022,Vendromin_2022}.
Consequently, these models are not valid for the pulsed multimode squeezed light in the high-gain regime.
In contrast, models with discrete losses, in which virtual beamsplitters are added between lossless media, have been shown to provide a simple and powerful approach for both the low-gain~\cite{Antonosyan_2014,Gr_fe_2017} and high-gain~\cite{Helt_2020,Quesada_2022} regimes.
However, the main disadvantage of this approach is the decrease in computational stability with increasing gain and losses. 
Thus, there is a huge demand for alternative approaches and numerical schemes.
 
The state of generated PDC light in a lossy medium is not pure, therefore the BMr is invalid and the basis of broadband Schmidt modes is not well defined.
The problem of finding the proper basis is crucial for applications and protocols that use frequency filtering of multimode squeezed light~\cite{Christ_2014,Melalkia_2022,Houde_2023}. 
Moreover, many continuous variable quantum information protocols~\cite{Weedbrook_2012} require high squeezing, so, similar to the lossless case, it is important to find a basis with the highest possible squeezing for the chosen lossy system. 

There are at least two well-known methods for finding the broadband modes which are valid for non-pure states. 
One of them is the so-called Mercer expansion, namely the diagonalization of the first-order coherence function. 
This approach was introduced by E.~Wolf and is widely used in classical coherence theory~\cite{Wolf_1982,Mandel_Wolf_book}.
In turn, the Williamson decomposition of the covariance matrix~\cite{Fabre_2020,Weedbrook_2012,Simon_1994} is a common method for the analysis of quantum Gaussian states.
Together with the Euler decomposition, it allows one to represent an arbitrary Gaussian state as a multimode thermal state amplified by an ideal multimode squeezer, what is widely used for squeezing extraction, see, e.g., Refs.~\cite{Quesada_2022,Hosaka_2016}. 
However, in a lossy system, neither the Mercer expansion, nor the Williamson-Euler decomposition guarantees that the maximal squeezing can be achieved in these bases.

In this paper, we present a unified theoretical approach to describe multimode squeezed light generation in lossy media.
Our approach includes both a discrete loss model based on a set of beamsplitters and a continuous loss model based on the spatial Langevin equation. 
For Gaussian input states and a non-displaced Markovian Gaussian environment, we derive master equations for the second-order correlation functions. 
To introduce broadband modes of lossy PDC, we consider the Mercer expansion and the Williamson-Euler decomposition and demonstrate that these bases are not optimal for squeezing measurement. 
In addition, we present an algorithm for the construction of a basis in which the maximal squeezing for multimode PDC can be reached. 
The properties of different broadband bases are demonstrated and discussed.

This paper is organized as follows. 
In Sections~\ref{subsec_gaussian},~\ref{sec_pdc_lossless} we define the relevant quantities, consider the main properties of Gaussian states and present a description of lossless multimode PDC. 
Section~\ref{sec_pdc_descrete_losses} presents a discrete model of losses based on the beamsplitter approach, where the solution is found using the transfer matrix method for the Bogoliubov transformation. 
In Section~\ref{sec_pdc_continuous_losses}  a continuous model of losses based on the Langevin equation is presented and the master equations for second-order correlation matrices are derived.
In Section~\ref{sec_decompositions} we consider two well-known algorithms to determine broadband modes of squeezed light (the Mercer expansion and the  Williamson-Euler decomposition) and compare them with the developed here algorithm which allows us to find the maximally-squeezed modes. 
Section~\ref{sec_results} contains the results of numerical simulations of PDC in a lossy medium, where both frequency-dependent and frequency-independent losses are considered.  
We conclude with Section~\ref{sec_conclusions}.
In addition, several theoretical and technical details are provided in appendices.

\section{Theoretical approach}
\label{sec_approach}

\subsection{Multimode Gaussian states}
\label{subsec_gaussian}

Without loss of generality, we consider a discrete frequency space, which allows us to make the intermediate derivations more transparent and simplify the routine of numerical calculations. To describe the frequency multimode quantum electromagnetic field at position $z$, we use the vector of annihilation operators
$\hat{\mathbf{a}}(z) = (\hat{a}_0, \hat{a}_1, \dots, \hat{a}_N)^T$, where the subscript defines the frequency mode, namely, $\hat{a}_i \equiv \hat{a}(z,\omega_i)$. 
These operators obey the bosonic commutation relations $[\hat{a}(z,\omega_i), \hat{a}^\dagger(z,\omega_j)]=\delta_{ij} $. 

It is well-known~\cite{Weedbrook_2012} that the Gaussian states can be fully described by the vector of the first-order moments $\braket{\hat{\mathbf{a}}}$ with elements $\braket{\hat{a}_i}$ and the second-order correlation matrices 
$\braket{\hat{\mathbf{a}}^\dagger \hat{\mathbf{a}}}$ and $\braket{\hat{\mathbf{a}} \hat{\mathbf{a}}}$ with matrix elements $\braket{\hat{a}_i^\dagger \hat{a}_j}$ and $\braket{\hat{a}_i \hat{a}_j}$, respectively.
Note that the matrix $\braket{\hat{\mathbf{a}}^\dagger \hat{\mathbf{a}}}$ is Hermitian, while the matrix $\braket{\hat{\mathbf{a}} \hat{\mathbf{a}}}$ is complex and symmetric.

For each frequency mode, one can introduce the dimensionless quadrature operators  $\hat{q}_i = \hat{a}^\dagger_i + \hat{a}_i $ and $\hat{p}_i =  i(\hat{a}^\dagger_i - \hat{a}_i) $ with the commutation relations $[\hat{q}_n, \hat{p}_m] = 2i\delta_{nm}$.
These operators define the covariance matrix $\sigma^{a}$, namely, the real positive-definite symmetric matrix of the second-order moments of the quadrature operators~\cite{Weedbrook_2012,Simon_1994}. 
For Gaussian states, the elements of the matrix $\sigma^{a}$ are given by
\begin{equation}
  \sigma^{a}_{ij} = \dfrac{\braket{\hat{x}_i\hat{x}_j + \hat{x}_j\hat{x}_i }}{2}-\braket{\hat{x}_i}\braket{\hat{x}_j},
\end{equation}
where $\hat{x}_i$ are the elements of the vector $\hat{\mathbf{x}}  = (\hat{q}_1, \hat{q}_2, \dots, \hat{q}_N, \hat{p}_1, \hat{p}_2, \dots, \hat{p}_N)^T$.
The covariance matrix includes the following set of expectation values: 
\begin{align}
    \braket{ \hat{q}_i \hat{q}_j} = 
    \delta_{ij} 
     + 2 \Big(\text{Re}[\braket{\hat{a}^\dagger_i \hat{a}_j}]
     + \text{Re}[\braket{\hat{a}_i \hat{a}_j}] \Big), \\
    \braket{ \hat{p}_i \hat{p}_j}  = 
    \delta_{ij} 
     + 2 \Big(\text{Re}[\braket{\hat{a}^\dagger_i \hat{a}_j}]
     - \text{Re}[\braket{\hat{a}_i \hat{a}_j}] \Big), \\
    \dfrac{ \braket{ \hat{p}_i \hat{q}_j +  \hat{q}_j \hat{p}_i} }{2} =  2 \Big(\text{Im}[\braket{\hat{a}_i \hat{a}_j}] -\text{Im}[\braket{\hat{a}^\dagger_i \hat{a}_j}]   
       \Big), \\
       \braket{\hat{q}_i}=2\text{Re}[\braket{\hat{a}_i}], ~ ~ ~
       \braket{\hat{p}_j}=2\text{Im}[\braket{\hat{a}_j}].
\end{align}

The diagonal elements of the covariance matrix correspond to the quadrature variances $(\Delta q_n)^2$ and $(\Delta p_n)^2$.
Note that for the vacuum state the covariance matrix is diagonal with $(\Delta q)^2 = (\Delta p)^2=1$. 

Using a unitary transformation $U$, one can move from the monochromatic basis $\hat{\mathbf{a}}$ to a broadband basis  $\hat{\mathbf{A}}= U \hat{\mathbf{a}}$.
In this notation, the rows of unitary matrix $U$ correspond to a new set of orthogonal modes.
The correlation matrices $\braket{\hat{\mathbf{A}}^\dagger \hat{\mathbf{A}}}$ and $\braket{\hat{\mathbf{A}} \hat{\mathbf{A}}}$ in the broadband basis can be obtained using the following transformations
 \begin{equation}
    \braket{\hat{\mathbf{A}}^\dagger \hat{\mathbf{A}}} = U^* \braket{\hat{\mathbf{a}}^\dagger \hat{\mathbf{a}}} U^T, ~ ~ \braket{\hat{\mathbf{A}} \hat{\mathbf{A}}} = U \braket{\hat{\mathbf{a}} \hat{\mathbf{a}}} U^T,
    \label{eq_correlation_transformation}
\end{equation}
where the superscripts $[.]^T$ and $[.]^*$ denote the matrix transpose and  the scalar complex conjugate, respectively. 
Further in the text, the conjugate transpose (Hermitian conjugate) of matrices is denoted as $[.]^H$.

In addition, one can introduce the broadband quadrature operators $\hat{Q}_n = \hat{A}^\dagger_n + \hat{A}_n $ and $\hat{P}_n =  i(\hat{A}^\dagger_n - \hat{A}_n) $, where the subscript $n$ denotes the mode index.
The covariance matrix in the broadband basis $\sigma^{A}$ is connected to the monochromatic covariance matrix $\sigma^{a}$ via the following transformation 
\begin{equation}
   \sigma^{A} = O \sigma^{a} O^T,
    \label{eq_covariance_matrix}
\end{equation} 
where the matrix
\begin{equation}
  O = \begin{bmatrix}
        \mathrm{Re}(U) &  -\mathrm{Im}(U) \\
        \mathrm{Im}(U) &  ~\mathrm{Re}(U) 
      \end{bmatrix}
      \label{eq_symplectic_transform}
\end{equation}
is orthogonal and symplectic.

Note that the requirement for the matrix $U$ to be unitary is equivalent to requiring the matrix $O$ to be orthogonal and symplectic: Under these properties, the commutation relations for the annihilation/creation and quadrature operators are preserved. 

\subsection{PDC generation without losses}
\label{sec_pdc_lossless}

In this work, we focus on the type-I PDC process, however, all the presented results can be easily extended to other types of PDC and FWM processes. 
A detailed derivation of the spatial Heisenberg equation~\cite{Huttner_1990,Horoshko_2022} for the fast-varying operators is presented in Appendix~\ref{appendix_lossless_pdc} and has the form
\begin{equation}
 \dfrac{d \hat{a}(z, \omega_i)  }{dz}  = i k_i \hat{a}(z, \omega_i) + i\Gamma \sum_{j} J_{ij}(z) \hat{a}^\dagger(z, \omega_j),
 \label{eq_heisenberg_equation}
\end{equation} 
where $\Gamma \propto \chi^{(2)} E_p$ is a real coupling constant and $k_i \equiv k(\omega_i) = n(\omega_i)\omega_i/c$.
The $z$-dependent coupling matrix is defined as
\begin{equation}
  J_{ij}(z) = S(\omega_i+\omega_j)e^{i k_p(\omega_i+\omega_j) z},
  \label{eq_coupling_matrix}
\end{equation}
where $S(\omega)$ is the initial pump spectrum, $k_p(\omega)=n_p(\omega)\omega/c$, while $n_p(\omega)$ and $n(\omega_i)$ are refractive indices of the pump and PDC photons, respectively.
Note, that the coupling matrix $J_{ij}(z)$ explicitly depends on $z$.
This means that the spatial-ordering effects are fully taken into account in Eq.~\eqref{eq_heisenberg_equation}.

The solution to the spatial Heisenberg equation~\eqref{eq_heisenberg_equation} in a nonlinear medium of length $L$ has the form of the Bogoliubov transformation
 \begin{equation}
        \hat{\mathbf{a}}(L) =  E \ \hat{\mathbf{a}}(0) +  F \ \hat{\mathbf{a}}^\dagger(0) ,
        \label{eq_bogoliubov_full}
 \end{equation}
where $ \hat{\mathbf{a}}(L)$ and $ \hat{\mathbf{a}}(0)$  are the vectors of the input ($z=0$) and  output ($z=L$) annihilation operators, respectively.
The matrices $E$ and $F$ are the Bogoliubov transfer matrices defined by the differential equations (see Appendix~\ref{appendix_lossless_pdc}, Eqs.~\eqref{eq_diff_Efunc} and \eqref{eq_diff_Ffunc}).

One of the most native ways to analyze the solution in the form of the Bogoliubov transformation is to perform the Bloch-Messiah reduction (BMr)~\cite{Braunstein_2005}.
BMr states that there exists a simultaneous singular value decomposition (SVD) of the Bogoliubov transfer matrices $E$ and $F$ in the following form (for details see Appendix~\ref{appendix_bogoliubov_full}):
\begin{align}
        E =  ~ \mathcal U^H \Lambda_E \mathcal W_E ~  ~  \text{and} ~ ~
        F =  ~ \mathcal U^H \Lambda_F  ( \mathcal W_E)^*,
        \label{eq_bloch_messiah_reduction}
\end{align}
where the matrices $\mathcal U$ and $\mathcal W_E$ are unitary and the matrices $\Lambda_E$ and $\Lambda_F$ are real and diagonal. 

The rows of matrices $\mathcal W_E$ and $\mathcal U$ correspond to the broadband modes at the input and output of the nonlinear medium, respectively.
The new input $\hat{\mathbf{A}}(0) =  \mathcal W_E \hat{\mathbf{a}}(0)$ and output $\hat{\mathbf{A}}(L) = \mathcal U  \hat{\mathbf{a}}(L)$ annihilation operators obey the bosonic commutation relations $[\hat{A}_i(z), \hat{A}^\dagger_j(z)]=\delta_{ij}$, $[\hat{A}_i(z), \hat{A}_j(z)]=0$.
In such a broadband basis,  the Bogoliubov transformation~\eqref{eq_bogoliubov_full} has a diagonal form
\begin{equation}
   \hat{\mathbf{A}}(L) = \Lambda_E \hat{\mathbf{A}}(0) + \Lambda_F \hat{\mathbf{A}}^\dagger(0).
\end{equation}

In what follows, the basis defined by the BMr, namely $U_{\mathcal{S}} \equiv (\mathcal U^{H})^{-1} = \mathcal U $, is called the Schmidt-mode basis. 

For a vacuum input, PDC results in a \textit{squeezed vacuum state}, a pure non-displaced Gaussian state.
Modes of squeezed vacuum (Schmidt modes) obtained with the use of BMr have the following important properties.

Firstly, there is an absence of field correlations between different Schmidt modes, namely, $\braket{\hat{A}_i^\dagger(L) \hat{A}_j(L)} = \lambda_{i,F}^2  \delta_{ij}$ and $\braket{\hat{A}_i(L) \hat{A}_j(L)} = \lambda_{i,E} \lambda_{i,F} \delta_{ij}$, where $\lambda_{i,E}$ and $\lambda_{i,F}$ are the elements of diagonal matrices  $\Lambda_E$ and $\Lambda_F$, respectively. 
Consequently, in the Schmidt-mode basis, the covariance matrix $\sigma^{\mathcal{S}}$ is diagonal and the output state can be interpreted as a set of independent broadband squeezers.
In order to measure the maximal degree of squeezing for a multimode squeezed vacuum state, the measurement must be carried out in the first Schmidt mode.

Secondly, BMr minimizes the number of modes describing the output state.
The number of occupied (amplified) Schmidt modes $K_\mathcal{S}$ is  given by
\begin{equation}
    K_{\mathcal{S}} = \Big( \sum_i n_i^2 \Big)^{-1},
    \label{eq_schmidt_number}
\end{equation}
where $n_i=N_i/(\sum_i N_i)$ and $N_i=\braket{\hat{A}_i^\dagger(L) \hat{A}_i(L)}$ are the normalized and absolute number of photons in mode $i$, respectively.
$K_\mathcal{S}$ is also known as the Schmidt number.
Because the BMr diagonalizes the matrix $\braket{\hat{A}_i^\dagger(L) \hat{A}_j(L)} = \lambda_{i,F}^2  \delta_{ij}$, the number of occupied modes in the Schmidt basis is minimal compared to any other basis (for details see Appendix~\ref{appendix_prove_occ_modes}).

\subsection{Discrete model of losses: transfer matrix method for the Bogoliubov transformation}
\label{sec_pdc_descrete_losses}

\begin{figure}[h!]
      \includegraphics[width=0.98\linewidth]{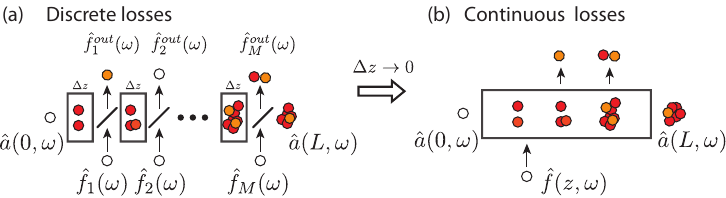}\\
    \caption{The PDC generation scheme in lossy media: (a) discrete model of losses (beamsplitter approach) and (b) continuous model of losses (Langevin equation).
                }
    \label{fig_scheme_pdc_losses}
\end{figure}

Here, we consider a lossy nonlinear medium of length $L$ with a frequency-dependent extinction coefficient $\alpha(\omega)$.
For simplicity, we assume that $\alpha(\omega)$ is constant along the whole nonlinear medium which results in  the light intensity transmission coefficient of the form $T(\omega)=e^{-\alpha(\omega)L}$.

In quantum optical systems,  losses are usually introduced by virtual beamsplitters.
To do this for PDC process, we cut the lossy nonlinear medium into $M$-segments of equal length  $\Delta z$ and assume that each segment consists of an ideal lossless squeezer followed by a beamsplitter, see Fig.~\ref{fig_scheme_pdc_losses}. 
In this case, the field transmission coefficient of each beamsplitter is given by $t(\omega)=e^{-\alpha(\omega)\Delta z/2}$ and corresponds to the amount of energy lost by a single segment of length $\Delta z$.
Each beamsplitter removes part of the initial PDC radiation and adds to the generated light a new uncorrelated field from the environment. 
This field is described by the vector of operators $\hat{\mathbf{f}}_{m}$ whose components obey the commutation relation $[\hat{f}_{in},\hat{f}^\dagger_{jm}]=\delta_{ij}\delta_{nm}$, where the first indices ($i, j$) correspond to the frequencies $\omega_{i}$ and $\omega_{j}$, while the second ones ($n, m$) correspond to the spatial coordinates $z_{n}$ and $z_{m}$.
Note that so far, as each mode of the environment interacts with the system only once, this model corresponds to the Markovian approximation~\cite{Vogel_Welsch_book,Gardiner_Zoller_book}.

\begin{figure}[h!]
      \includegraphics[width=0.89\linewidth]{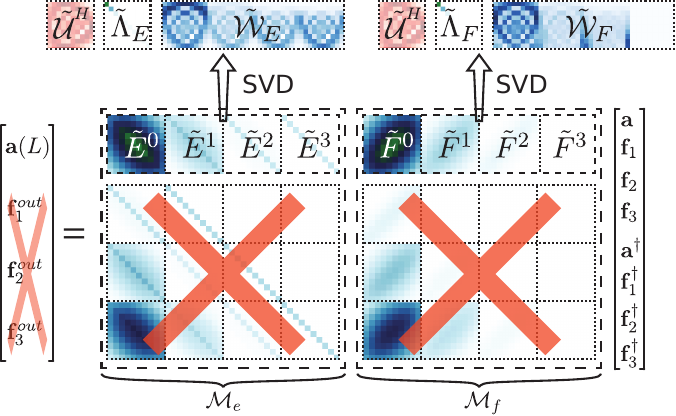}\\
    \caption{Visualization of the Bogoliubov transformation for a discrete model with three virtual beamsplitters ($M=3$) dividing the nonlinear medium into three segments. 
    The left vector corresponds to the set of output annihilation operators, the right vector represents the set of input operators.
    The matrices $\mathcal{M}_e$ and $\mathcal{M}_f$ describe the full Bogoliubov transformation. 
    According to  Eq.~\eqref{eq_partial_bogoliubov}, the PDC field operators at the output of the nonlinear medium $\hat{\mathbf{a}}(L)$ are determined only by the matrices $\tilde E^{m}$ and $\tilde F^{m}$ (partial Bogoliubov transformation), therefore, there is no need to calculate the full Bogoliubov transformation:
     the crossed matrices describing the evolution of the environment operators $\hat{\mathbf{f}}^{out}_m$ can be ignored. The top row visualizes the SVD of the matrices $\tilde E^{m}$ and $\tilde F^{m}$ according to Eq.~\eqref{eq_SVD_partial}. 
     The red color corresponds to the unitary matrix $\tilde{\mathcal U}^H$ that defines the broadband basis of the output operators, while the blue color corresponds to the matrices $\tilde{\mathcal W}_e$ and $\tilde{\mathcal W}_f$ that define two broadband bases of the input operators. 
     Each small square block has $N\times N$ size, where $N$ is the number of monochromatic modes.
                }
    \label{fig_scheme_bogoliubov}
\end{figure}

In general, the obtained system with the all input/output environment modes implies a unitary evolution and, therefore, there exists the Bogoliubov transformation that represents a unique map between the input and output modes, see Fig.~\ref{fig_scheme_bogoliubov}.  
In what follows, this Bogoliubov transformation is called the full Bogoliubov transformation.
The full Bogoliubov transformation can be found by multiplying the set of local Bogoliubov transformations of each segment $\Delta z$ (see Appendix~\ref{appendix_discrete_model}).

The number of matrix elements of the full Bogoliubov transformation scales as  $(N (M+1))^2$, where $M$ is the number of virtual beamsplitters and $N$ is the number of monochromatic frequency modes.
The full Bogoliubov transformation gives both the output modes $\hat{\mathbf{a}}(L)$ and the environment modes $\hat{\mathbf{f}}^{out}_n$.
However, since the only useful and measurable information is contained in the output modes $\hat{\mathbf{a}}(L)$, it is sufficient to compute the corresponding sub-matrices $\tilde E^{m}$ and $\tilde F^{m}$, see Fig.~\ref{fig_scheme_bogoliubov}, instead of calculating the full Bogoliubov transformation, which greatly simplifies the numerical effort. 
Below, we refer to these sub-matrices as the partial Bogoliubov transformations.

The explicit expressions of  partial Bogoliubov transformation matrices $\tilde E^{m}$ and $\tilde F^{m}$ can be derived using the Bogoliubov transform matrix method and are presented in detail in Appendix~\ref{appendix_discrete_model}.
In such a lossy model, the output annihilation operator of the generated PDC light has the following form 
\begin{equation}
        \hat{\mathbf{a}}(L) =\tilde E^{0} \hat{\mathbf{a}}(0) +  \tilde F^{0} \hat{\mathbf{a}}^\dagger(0)   + \sum_{m=1}^M \Big[ \tilde E^{m} \hat{\mathbf{f}}_{m} + \tilde F^{m} \hat{\mathbf{f}}^\dagger_{m}  \Big].
        \label{eq_partial_bogoliubov}
 \end{equation}

In contrast to the lossless solution, the output annihilation operators $\hat{\mathbf{a}}(L)$ in a lossy medium contain additional terms: 
the nonzero matrices $\tilde F^{m}$  demonstrate that not only the initial field $\hat{\mathbf{a}}^\dagger(0)$, but also the environment fields $\hat{\mathbf{f}}_m^\dagger$ are amplified.

Moreover, unlike for the full Bogoliubov transformation, the partial Bogoliubov transformation describes an open system and is therefore not invertible, so the BMr cannot be applied in this case (see Appendix~\ref{appendix_bogoliubov_part} for details). 
However, for the partial Bogoliubov transformation defined by the matrices 
\begin{equation}
    \tilde{\mathbf{E}} = (  \tilde E^{0} \:  \tilde E^{1} \:  \tilde E^{2} \dots   \tilde E^{M}),  ~ ~ \tilde{\mathbf{F}} = (  \tilde F^{0} \:  \tilde F^{1} \:  \tilde F^{2} \dots   \tilde F^{M}),
\end{equation}  
 there exists a simultaneous SVD in the form
\begin{equation}
  \tilde{\mathbf{E}} = \tilde{\mathcal U}^H \tilde{\Lambda}_E \tilde{\mathcal W}_E, ~ ~
  \tilde{\mathbf{F}} = \tilde{\mathcal U}^H \tilde{\Lambda}_F \tilde{\mathcal W}_F.
  \label{eq_SVD_partial}
\end{equation}
Here, the left unitary matrix $\tilde{\mathcal U}^H$ (of size  $N\times N$) is the same for both decompositions, 
and allows us to introduce the orthogonal basis of modes $U_{M} \equiv (\tilde{\mathcal U}^{H})^{-1} = \tilde{\mathcal U}$, which defines the output broadband operators $\hat{\mathbf{A}}(L) =  U_{M}  \hat{\mathbf{a}}(L)$. 

The matrices $\tilde{\mathcal W}_E$ and $\tilde{\mathcal W}_F$ are $N\times N(M+1)$ matrices.
The rows of matrix $\tilde{\mathcal W}_E$  are orthogonal, so $\tilde{\mathcal W}_E (\tilde{\mathcal W}^*_E)^T=\mathbf{1}_N$ (same for  $\tilde{\mathcal W}_F$: $\tilde{\mathcal W}_F (\tilde{\mathcal W}^*_F)^T=\mathbf{1}_N$), where $\mathbf{1}_N$ is the identity matrix of size $N\times N$.
In a similar way, the annihilation operators can be introduced with the matrices  $\tilde{\mathcal W}_E$ and $\tilde{\mathcal W}_F$. 
So far as the matrices $\tilde{\mathcal W}_E$ and $\tilde{\mathcal W}_F$ are different ($\tilde{\mathcal W}_E (\tilde{\mathcal W}^*_F)^T \neq \mathbf{1}_N$), one should introduce two different sets of input broadband operators: $\hat{\mathbf{B}}_E =  \tilde{\mathcal W}_E \: \mathrm{col}[\hat{\mathbf{a}}(0), \hat{\mathbf{f}}_1,  \hat{\mathbf{f}}_2, ... \hat{\mathbf{f}}_N ]$ and $\hat{\mathbf{B}}^\dagger_F =  \tilde{\mathcal W}_F \: \mathrm{col}[\hat{\mathbf{a}}^\dagger(0), \hat{\mathbf{f}}^\dagger_1,  \hat{\mathbf{f}}^\dagger_2, ... \hat{\mathbf{f}}^\dagger_N ] $, where the $\mathrm{col}(...)$ means that the vectors of operators are stacked column-wise.
The commutation relation for each set of operators is bosonic, however,
 $[\hat{B}_{E, n}, \hat{B}^{\dagger}_{F, m}]\neq \delta_{nm}$.

By analogy with the lossless case, the introduced broadband operators allow us to write the input-output relation~\eqref{eq_partial_bogoliubov} in a diagonal form 
\begin{equation}
   \hat{\mathbf{A}}(L) = \tilde\Lambda_E \hat{\mathbf{B}}_E + \tilde \Lambda_F \hat{\mathbf{B}}^\dagger_F.
   \label{eq_solution_losses}
\end{equation} 

Let us consider the initial vacuum state and vacuum environment.
Using Eq.~\eqref{eq_solution_losses}, one can notice that the matrix $\braket{\hat{A}_{i}^\dagger(L) \hat{A}_{j}(L)} = \tilde\lambda_{i,F}^2 \delta_{ij}$ becomes diagonal which corresponds to the so-called Mercer expansion~\cite{Raymer_2020,Wolf_1982}.
Further in the text, the basis of modes $U_{M}$ will be called the Mercer-Wolf basis.

At the same time, the correlation matrix  $\braket{\hat{A}_{i}(L) \hat{A}_{j}(L)}$ is not diagonal due to $\tilde{W}_E (\tilde{W}^*_F)^T \neq \mathbf{1}_N$.
As a result,  in the Mercer-Wolf basis, the covariance matrix contains non-diagonal terms. 

The SVD-decomposition of the partial Bogoliubov transformation is performed for $N\times N(M+1)$ matrices. 
This is quite costly for large $M$, so in Section~\ref{subsubsec_mercer} an alternative method for the computation of the Mercer-Wolf modes is presented. 

\subsection{Continuous model of losses: Langevin and master equations  }
\label{sec_pdc_continuous_losses}

\subsubsection{ Langevin equation  }
\label{subsubsec_Langevin}

In order to obtain a model that is able to describe losses continuously (see Fig.~\ref{fig_scheme_pdc_losses}(b)), we consider a single-segment solution of the discrete model, given by Eq.~\eqref{eq_partial_bogoliubov}, in the interval $[z_0, z]$, where $z=z_0+\Delta z $. 
An explicit expression of such a solution is derived in Appendix~\ref{appendix_discrete_model}  (see Eq.~\eqref{eq_app_segment_solution}) and has the form  
\begin{multline}
    \hat{a}_i(z) = t_i(\Delta z) \Big( \sum_{n} E_{in}(z_0+\Delta z, z_0)\hat{a}_n(z_0) \\ + F_{in}(z_0+\Delta z, z_0)\hat{a}^\dagger_n(z_0) \Big) + r_i(\Delta z)  \hat{f}_i ,
    \label{eq_segment_solution}
\end{multline}
where $t_i(\Delta z)  = e^{-\frac{\alpha_i}{2} \Delta z} $ and $   r_i(\Delta z)  = \sqrt{1-e^{-\alpha_i \Delta z}}$ are the transmission and reflection coefficients of the beamsplitter, respectively.

For small $\Delta z$, the functions $t_i(\Delta z)$ and $r_i(\Delta z)$ can be expanded using the Taylor series: 
\begin{align}
  t_i(\Delta z) &  \approx 1-\frac{\alpha_i \Delta z}{2}+O(\Delta z^2), 
  \label{eq_limit_dz_1}
  \\
      r_i(\Delta z) & \approx \sqrt{\alpha_i } \sqrt{\Delta z} + O(\Delta z^{3/2}).
      \label{eq_limit_dz}
\end{align}

Considering the operator derivative in the form $ \frac{d \hat{a}_i }{d z} \equiv \lim_{\Delta z \rightarrow0} [\hat{a}_i(z+\Delta z)-\hat{a}_i(z)]/{\Delta z}$ and taking into account Eqs.~\eqref{eq_limit_dz_1}, \eqref{eq_limit_dz} and the Heisenberg equation~\eqref{eq_heisenberg_equation}, one can obtain a differential equation for the lossy operators Eq.~\eqref{eq_segment_solution} in the form of the Langevin equation:
\begin{multline}
    \dfrac{d \hat{a}_i(z) }{d z} = i \kappa_i \hat{a}_i(z)  \\ + i\Gamma \sum_{s} J_{is}(z) \hat{a}^\dagger_s(z)  + \sqrt{\alpha_i } \hat{f}_i(z),   
    \label{eq_langevin}
\end{multline}
where $J_{is}(z)$ is the coupling matrix (same as in Eq.~\eqref{eq_coupling_matrix}) and $\Gamma$ is the coupling coefficient. 
Here, the quantity $\kappa_i = k_i+i\alpha_i/2$ can be interpreted as a complex wave-vector in a lossy medium for the frequency $\omega_i$.
The continuous Langevin noise operator $ \hat{f}(z, \omega_i) \equiv  \hat{f}_i(z) $ is introduced as~\cite{Vogel_Welsch_book} $ \hat{f}_i(z) \equiv \lim_{\Delta z \rightarrow0} \frac{ \hat{f}_i }{\sqrt{\Delta z}}$ 
and obeys the commutation relation $[\hat{f}(z, \omega_i), \hat{f}^\dagger(z^\prime, \omega_j)] = \delta_{ij}\delta(z-z^\prime)$.
 
Oppositely to cavity PDC, where systems have stationary dynamics on a large time scale and are described by the steady-state solution~\cite{Gardiner_Zoller_book,Gouzien_2020,Onodera_2022,Guidry_2023},
we consider parametric amplification in a crystal of finite length, where the dynamics is defined by the $z$-dependent coupling matrix $J_{is}(z)$ and is non-stationary.
Therefore, for the studied regime, we have to solve the Langevin equation explicitly.

The Langevin equation, in which the complex wave-vector is independent on $z$, has the form of a non-autonomous and non-homogeneous differential equation with the solution of the form~\cite{Agarwal_2008}:
\begin{multline}
    \hat{a}_i(z) = \sum_j \Big( g_{ij}(z, z_0) \hat{a}_j( z_0) + h_{ij}(z, z_0) \hat{a}^\dagger_j( z_0) \Big) \\ +   \sqrt{\alpha_i } \int_{z_0}^z \: dx \sum_j \Big( g_{ij}(z, x)\hat{f}_j(x) + h_{ij}(z, x) \hat{f}_j^\dagger(x) \Big).
    \label{eq_cont_langevin}
\end{multline}

Here, the functions $g_{ij}(z, x)$ and $h_{ij}(z, x)$ are determined by the following
differential equations: 
\begin{align}
    \partial_z g_{ij}(z, x)  =  i\kappa_i g_{ij}(z, x) + i\Gamma \sum_{s} J_{is}(z) h^*_{sj}(z, x) ,
    \label{eq_func_h} \\
    \partial_z h_{ij}(z, x)  = i\kappa_i h_{ij}(z, x) +  i\Gamma \sum_{s} J_{is}(z)  g^{*}_{sj}(z, x) 
    \label{eq_func_g}
\end{align}
with the initial conditions $g_{ij}(x, x)=\delta_{ij}$ and $h_{ij}(x, x)=0$.

Note that the partial Bogoliubov transformation~\eqref{eq_partial_bogoliubov} is a discretized version of the continuous solution~\eqref{eq_cont_langevin}.

\subsubsection{ Master-equations for Gaussian states  } 
\label{subsubsec_Master}

The Langevin equation in the form Eq.~\eqref{eq_langevin} is valid for any input state.
However, for Gaussian input states, the output state is completely determined by moments up to the second order. 
If, in addition, the output state is not displaced, the first-order moments vanish and the state is fully characterized by only the second-order moments (correlation matrices) $\braket{\hat{a}^\dagger_i(z) \hat{a}_j(z)}$ and $\braket{\hat{a}_i(z) \hat{a}_j(z)}$. 

To obtain master-equations for the correlation matrices, the solution to the Langevin equation Eq.~\eqref{eq_cont_langevin} together with Eqs.~\eqref{eq_func_h} and \eqref{eq_func_g} are used.
 The derivative of the operator $\hat{a}^\dagger_i(z) \hat{a}_j(z)$ can be either calculated explicitly or considered in a limiting form:
\begin{equation}
  \frac{d \braket{\hat{a}^\dagger_i(z) \hat{a}_j(z)} }{d z} \equiv \lim_{\Delta z \rightarrow0}   \dfrac{\braket{\hat{a}^\dagger_i(z^\prime) \hat{a}_j(z^\prime) - \hat{a}^\dagger_i(z) \hat{a}_j(z)}}{\Delta z},
      \label{eq_derivative_correl}
\end{equation}
where $z^\prime=z+\Delta z$ (and analogously for $\braket{\hat{a}_i(z)\hat{a}_j(z)}$).

In this work, we assume a non-displaced ($\braket{\hat{f}_i^\dagger(z)} = \braket{\hat{f}_i(z)} = 0$) and  spatially delta-correlated (Markovian) environment: $\braket{\hat{f}^\dagger_i(z) \hat{f}_j(z^\prime)} = \braket{\hat{f}^\dagger_i(z) \hat{f}_j(z)}\delta(z-z^\prime)$ and $\braket{\hat{f}_i(z) \hat{f}_j(z^\prime)} = \braket{\hat{f}_i(z) \hat{f}_j(z)}\delta(z-z^\prime)$. 
The frequency correlations $\braket{\hat{f}_i^\dagger(z)  \hat{f}_j(z) }$ and $\braket{\hat{f}_i (z)  \hat{f}_j(z) }$ are determined by the state of the environment at position $z$.
For example, for a spectrally uncorrelated \textit{thermal} environment, the spectral correlations read $\braket{\hat{f}_i^\dagger(z)  \hat{f}_j(z) } = \bar{N}^{e} \delta_{i, j}$ and $\braket{\hat{f}_i (z)  \hat{f}_j(z) } = 0$, where $\bar{N}^{e}_i$ is the average number of thermal photons. 
If the environment is in the vacuum state, then $\bar{N}^{e}_i=0$.
Note that the environment can be quite exotic: e.g. having  the finite correlation function $\braket{\hat{f}^\dagger(z, \omega)  \hat{f}(z, \omega^\prime) } = C(\omega, \omega^\prime)$  or $\braket{\hat{f}_i (z)  \hat{f}_j(z) } \neq 0$ as for the squeezed thermal bath~\cite{Paris_2003,Serafini_2004}.

The resulting master equations for correlation matrices read 
\begin{multline}
\dfrac{d \braket{\hat{a}^\dagger_i(z) \hat{a}_j(z)}}{d z} = 
        ( - i \kappa^*_{i} + i \kappa_{j} )  \braket{\hat{a}^\dagger_i(z)\hat{a}_j(z)}  \\ + 
           i\Gamma \sum_{s} \Big( J_{js}(z)   \braket{\hat{a}^\dagger_i(z)\hat{a}^\dagger_s(z)} - J^*_{is}(z) \braket{\hat{a}_s(z) \hat{a}_j(z)}\Big) \\
        + \sqrt{\alpha_i  \alpha_j}  \braket{\hat{f}_i^\dagger(z) \hat{f}_j(z)},
        \label{eq_correlation_equation_1}
\end{multline}

\begin{multline}
    \dfrac{d \braket{\hat{a}_i(z) \hat{a}_j(z)}}{d z} = ( i \kappa_{i} + i \kappa_{j} ) \braket{\hat{a}_i(z)\hat{a}_j(z)} \\ + i\Gamma  \sum_{s} \Big(J_{js}(z) \braket{\hat{a}_i(z)\hat{a}^\dagger_s(z)} + J_{is}(z) \braket{\hat{a}^\dagger_s(z)\hat{a}_j(z) } \Big) \\ 
     + \sqrt{\alpha_i  \alpha_j}  \braket{\hat{f}_i(z) \hat{f}_j(z)}. 
    \label{eq_correlation_equation_2}
\end{multline}

The initial conditions for the correlation matrices are determined by the initial Gaussian state (e.g. vacuum, coherent, thermal, etc.). 
For example, for the spectrally uncorrelated thermal input state, the initial conditions read $\braket{\hat{a}^\dagger_n(0) \hat{a}_m(0)}= \bar{N}_n \delta_{n, m}$ and $\braket{\hat{a}_n(0) \hat{a}_m(0)}=0$, where  $\bar{N}_n $ is the mean number of photons in mode $n$.
For the vacuum input state $\bar{N}_n=0$.
The Eqs.~\eqref{eq_correlation_equation_1} and \eqref{eq_correlation_equation_2} also hold, if the temperatures of the input and the environment fields are different. 

Note that both the Langevin and the master equations contain the coupling matrix $J(z)$, which explicitly depends on $z$. 
This $z$-dependence reflects the fact that the spatial-ordering effects are fully included in the approach we present.

In opposite to the discrete model, the master equations~\eqref{eq_correlation_equation_1} and \eqref{eq_correlation_equation_2} constitute an explicit system of the first-order differential equations, which can be solved numerically with the usage of stable high-order numerical algorithms, e.g. with the Runge-Kutta methods~\cite{Butcher_2008}.
So far as in the main text we present all equations for discrete frequencies, in Appendix~\ref{appendix_continuous_freq}, the Langevin and the master equations are presented for the continuous counterparts.

\subsection{Broadband modes for multimode squeezed states}
\label{sec_decompositions}

In this section, we present three different mode bases that can be used for the mode analysis of lossy broadband squeezed light.

\subsubsection{ Mercer-Wolf modes }
\label{subsubsec_mercer}

The first basis contains the Mercer-Wolf modes, which can be found with the Mercer expansion~\cite{Wolf_1982,Mandel_Wolf_book}.
So far as the matrix $\braket{\hat{\mathbf{a}}^\dagger\hat{\mathbf{a}}}$ is Hermitian, its diagonalization can be performed using the eigendecomposition based on the unitary matrix~$V$:
    \begin{equation}
        \braket{\hat{\mathbf{a}}^\dagger\hat{\mathbf{a}}} = V \braket{\hat{\mathbf{B}}^\dagger\hat{\mathbf{B}}} V^H, 
      \end{equation}  
where the matrix $\braket{\hat{\mathbf{B}}^\dagger\hat{\mathbf{B}}}$ is real and diagonal and its diagonal values are sorted in descending order.
Then, the inverse transform is
    \begin{equation}
       \braket{\hat{\mathbf{B}}^\dagger\hat{\mathbf{B}}} = V^H  \braket{\hat{\mathbf{a}}^\dagger\hat{\mathbf{a}}} V.
       \label{eq_Mercer}
      \end{equation}  
Comparing the transformation in Eq.~\eqref{eq_Mercer} with Eq.~\eqref{eq_correlation_transformation},  
one can introduce a broadband basis  $\hat{\mathbf{B}} =  V^T\hat{\mathbf{a}}$.
It should be noted that there is a freedom in choosing the   $V^T$ matrix: The eigenvectors are determined up to an arbitrary phase. 
To resolve this issue, we can align the ellipses in the quadrature phase space for each mode separately, in order to obtain the maximal variance for the quadrature $\hat{Q}$ and the minimal one for $\hat{P}$.
This can be done by finding the phases $\phi_i = \mathrm{arg} \big( \braket{\hat{B}_i\hat{B}_i} \big)$ and constructing the unitary matrix $ Y  = \mathrm{diag} (e^{-\frac{i\phi_1}{2}}, e^{-\frac{i\phi_2}{2}},..., e^{-\frac{i\phi_N}{2}}) $.

Finally, the proper unitary transformation for the Mercer-Wolf basis reads $U_{M} = Y V^T $ and the Mercer-Wolf modes are defined as 
\begin{equation}
    \hat{\mathbf{A}}_{M}  = U_{M}  \hat{\mathbf{a}}.
\end{equation}

Note that the Mercer-Wolf basis is characterized by the minimal number of occupied modes, for more details see Appendix~\ref{appendix_prove_occ_modes}.

\subsubsection{ Williamson-Euler modes}
\label{subsubsec_Williamson}

The second way to find broadband modes is based on Williamson decomposition of covariance matrix~\cite{Houde_2023,Weedbrook_2012,Simon_1994}, namely, a symplectic decomposition of the real symmetric positive matrix: 
\begin{equation}
    \sigma^a = S D S^T,
\end{equation} 
where $S$ is the symplectic  $2N \times 2N$ matrix, and $D=\text{diag}(\nu_1,\dots,\nu_N; \nu_1,\dots,\nu_N)$ is a diagonal matrix.
The real and positive values $\nu_i$ are known as the symplectic spectrum.

The symplectic matrix $S$ can be additionally decomposed using the Euler decomposition~\cite{Fabre_2020,Houde_2023,Weedbrook_2012} 
\begin{equation}
    S = O_l \Lambda O_r,
\end{equation}
where $O_l$ and $O_r$ are both orthogonal and symplectic matrices and $\Lambda=\text{diag}(e^{r_1},\dots,e^{r_N};e^{-r_1},\dots,e^{-r_N})$ is a diagonal matrix 
with the values $r_i$  sorted in descending order.
Note that the Euler decomposition of the symplectic matrix is nothing more than the Bloch-Messiah reduction, written for quadratures instead of annihilation and creation operators.

Finally, the covariance matrix reads
\begin{equation}
    \sigma^a = O_l \Lambda O_r  D O_r^T \Lambda  O_l^T.
    \label{eq:Euler}
\end{equation}
Comparing Eq.~\eqref{eq:Euler} with Eq.~\eqref{eq_symplectic_transform}, we can extract the required unitary transformation $U^W$ as
\begin{equation}
    O_l^T  = 
     \begin{bmatrix}
        \mathrm{Re}(U^W) & -\mathrm{Im}(U^W) \\
        \mathrm{Im}(U^W) & ~\mathrm{Re}(U^W)
     \end{bmatrix},
     \label{eq_transformation_will}
\end{equation}
and introduce new broadband annihilation operators: 
\begin{equation}
    \hat{\mathbf{A}}_W = U_W \hat{\mathbf{a}}.
\end{equation}
We define the $U_W$ basis as the \textit{Williamson-Euler mode basis}.
Numerically, the Williamson-Euler decomposition can be found with the use of algorithms presented in, e.g. Refs.~\cite{Safranek_2018,Gupt2019,houde_2024}.

For pure states, all the symplectic values are $|\nu_i|=1$ and the matrix $S$ is both symplectic and orthogonal.
Consequently, $\sigma^a = \tilde{O}_l \Lambda^2 \tilde{O}_l^T$ that has the form of the eigendecomposition of the covariance matrix.
The unitary transformation, reconstructed with Eq.~\eqref{eq_transformation_will} from $\tilde{O}_l$, gives  simply the Schmidt mode basis.
The matrix $\Lambda^2$ is then the covariance matrix in the Schmidt-mode basis.

\subsubsection{ Maximally squeezed (MSq) modes }
\label{subsubsec_Eigendecomposition}

The third basis studied in this paper includes modes that reveal the maximal squeezing of the system.

To obtain such a basis, we start with the eigendecomposition of the covariance matrix: 
The real symmetric covariance matrix $\sigma_a$ can be decomposed using the eigendecomposition:
\begin{equation}
    \sigma^a = O  \Lambda  O^T,
\end{equation} 
where $\Lambda$ is a diagonal matrix with real and positive values and $O$ is an orthogonal matrix.

For pure states, the matrix $O$ is symplectic, so the eigendecomposition and the Williamson-Euler decomposition coincide. 
For mixed states, the matrix $O$ is not symplectic, i.e., the matrix $\tilde U$ built from $O^T$ as
\begin{equation}
    O^T =   \begin{bmatrix}
                \mathrm{Re}(\tilde U) & -\mathrm{Im}(\tilde U) \\
                \mathrm{Im}(\tilde U) & ~\mathrm{Re}(\tilde U)
            \end{bmatrix}
\end{equation}
is not unitary. 
In other words, the new set of modes can not be defined by simply utilizing the eigendecomposition of the covariance matrix.

However, to define the basis of modes with maximal squeezing, one can apply the following procedure. 
Among real and positive eigenvalues of the covariance matrix $\sigma^a$ let us choose the minimal eigenvalue $\lambda^{(1)}_{min}$ and the corresponding eigenvector $v_{min}^{(1)} = (c_1, c_2, \dots, c_N, d_1, d_2, \dots, d_N)$.
Since this vector is normalized, we define the $P$-quadrature operator as $\hat{P_1} = \sum_n (c_n \hat{q}_n + d_n \hat{p}_n)$. 
The $Q$-quadrature operator is then clearly defined as $\hat{Q_1} = \sum_n (d_n \hat{q}_n - c_n \hat{p}_n)$.
As a result, the corresponding broadband operator is introduced as $ \hat{A}_1 = \sum_n (d_n+i c_n) \hat{a}_n $. 

Having the fixed basis vector $\phi_1^{(1)} = \big((d_1+i c_1), (d_2+i c_2), \dots,  (d_N+i c_N)\big)$, one can use the Gram-Schmidt orthogonalization procedure to construct the orthogonal set of vectors forming a unitary matrix $U^{(1)}$.
In this matrix, the first vector corresponds to the mode with the quadrature variance $(\Delta {P_1})^2 = \lambda^{(1)}_{min}$. 
Note that since $\lambda^{(1)}_{min}$ is the minimum eigenvalue of the covariance matrix, it defines the maximal squeezing that can be measured in the system.
This result is known as the \textit{squeezing criteria} for multimode Gaussian states~\cite{Simon_1994}.

So far, as the vectors $ \{\phi_n^{(1)} : n = 2, ..., N  \}$ of the matrix $U^{(1)}$ are constructed using the  Gram-Schmidt orthogonalization, their squeezing properties are random.
Nevertheless, given the first vector $\phi_1^{(1)}$,  we can follow the same procedure as above  to determine the maximal squeezing in the subspace of the remaining vectors of the matrix $U^{(1)}$.
Namely, we compute the covariance matrix $\sigma^{(2)}$, which is determined by the vectors $ \{\phi_n^{(1)} : n = 2, ... N  \}$.
In the same way, the eigenvector $v_{min}^{(2)} =  (c^{(2)}_1, c^{(2)}_2, \dots, c^{(2)}_N, d^{(2)}_1, d^{(2)}_2, \dots, d^{(2)}_N)$ with the minimal eigenvalue $\lambda^{(2)}_{min}$ of $\sigma^{(2)}$ is found. 
Similarly, the corresponding mode $\phi_1^{(2)}=\big((d^{(2)}_1+i c^{(2)}_1), (d^{(2)}_2+i c^{(2)}_2), \dots,  (d^{(2)}_N+i c^{(2)}_N)\big)$ is introduced.

As a result, we have two fixed orthogonal broadband modes $\phi_1^{(1)}$ and $\phi_1^{(2)}$, and using the Gram-Schmidt orthogonalization procedure we can build the unitary matrix $U^{(2)}$.
The first vector of this matrix corresponds to the mode with the quadrature variance $(\Delta {P_1})^2 = \lambda^{(1)}_{min}$, while the second one -- to the mode with the quadrature variance $(\Delta {P_2})^2 = \lambda^{(2)}_{min}$.

Repeating this procedure for all the modes, we obtain the basis $U^{(N)}$, in which the first mode corresponds to the maximally possible squeezing in the system.
At the same time, each mode $m$ with $m>1$ gives the maximally possible squeezing that can be achieved in the rest of the system without the preceding $m-1$  modes.
Further in the text we call this basis \textit{the maximally squeezed (MSq) mode basis} and denote the corresponding unitary matrix as $U_{MSq} \equiv U^{(N)}$. 
The numerical realization of the aforementioned algorithm can be found in Ref.~\cite{kopylov_2025_code}.

The covariance matrix in the MSq basis $\sigma^{MSq}$ has the, so-called, \textit{the second canonical form} which is obtained via `phase-space rotations' as described in
Ref.~\cite{Simon_1994}.
A similar recursive procedure was introduced for the operational construction of a basis for experimental measurements~\cite{Opatrn__2002}. 

\subsection{ Additional useful relations }
\label{subsubsec_additional}

There are several aspects that should be mentioned. 
First, all decompositions provide modes whose ellipses are aligned in the phase space:  There are no correlations between $P$- and $Q$-quadratures of the same mode, i.e., $\braket{ \hat{P}_n \hat{Q}_n + \hat{Q}_n \hat{P}_n } = 0$ and the maximum quadrature variance corresponds to the $\hat{Q}$-component. 
According to our quadrature definitions, the vacuum variance is $(\Delta P^{vac}_i)^2=1$; therefore, the inequality 
 $(\Delta P_i)^2 < 1$ defines the squeezing in the $i$th mode.
In what follows, the quadrature variances are given in dB as $ 10 \log_{10} \big[(\Delta P)^2 \big] $.

Secondly, for each basis we can compute the number of occupied modes. 
We define the number of modes in the selected basis $U$ in a similar way to the lossless case Eq.~\eqref{eq_schmidt_number}:
\begin{equation}
    K_U = \Big( \sum_i n_i^2 \Big)^{-1},
    \label{eq_number_of_modes}
\end{equation}
where $n_i=N_i/(\sum_i N_i)$ and $N_i=\braket{\hat{A}_i^\dagger \hat{A}_i}$  are the normalized and absolute number of photons in mode $i$ of the basis $U$, respectively. 

\begin{figure}
      \includegraphics[width=1.\linewidth]{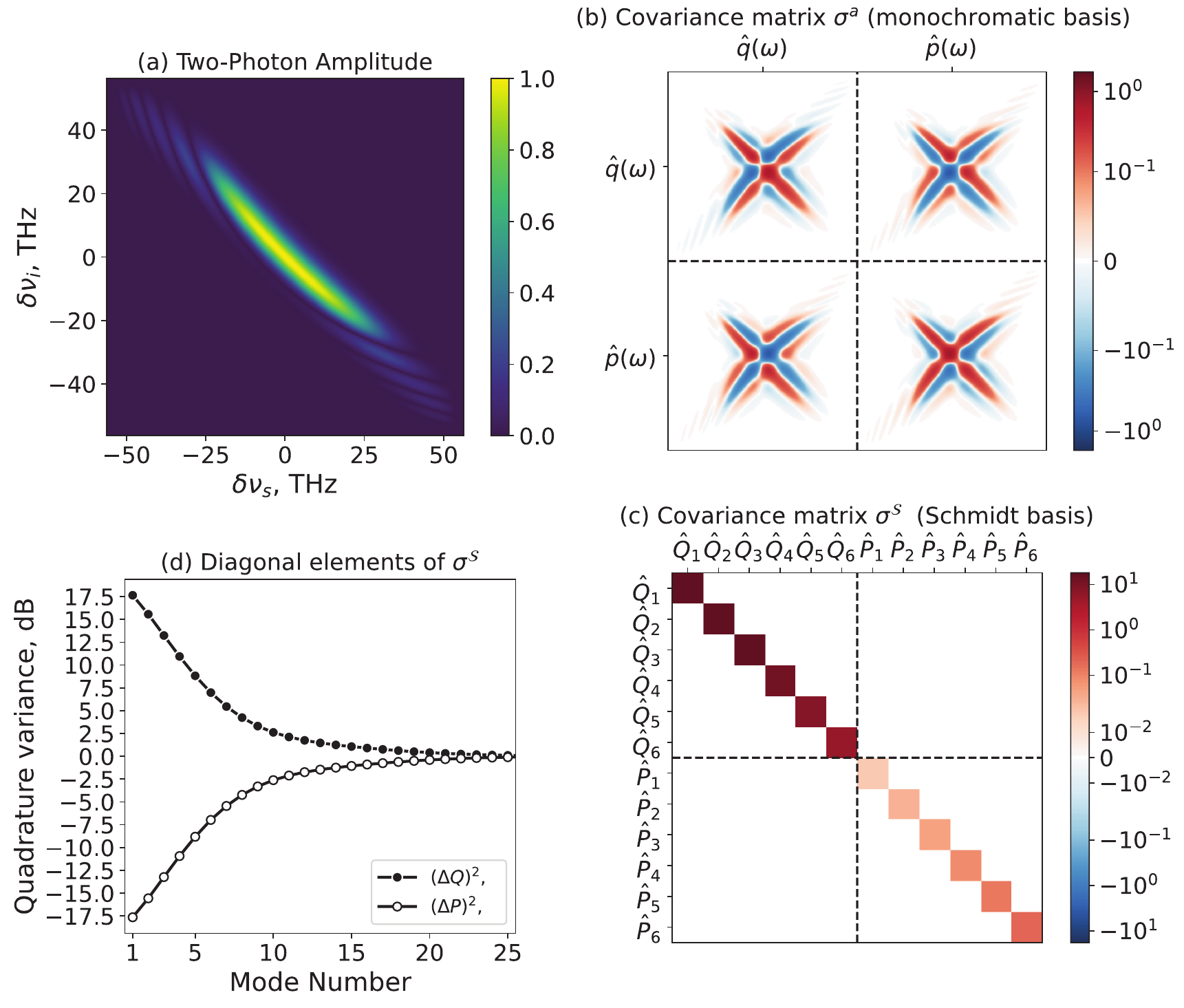 } 
    \caption{ Numerical results for the lossless PDC. 
    The parameters of the model are given in the main text.  
    (a) Two-photon amplitude $F(\omega_s,\omega_i)=S(\omega_i+\omega_j) \ \mathrm{sinc}(\Delta k(\omega_s,\omega_i) L /2)$ which characterizes PDC in a low-gain regime; $\delta \nu$ is the detuning from the central frequency of PDC $\nu_p/2$. 
    (b) and (c) Covariance matrices in the monochromatic and Schmidt-mode bases, respectively.
    (d) Quadrature variances for the first 25 Schmidt modes. Filled and empty markers correspond to $(\Delta {Q_m})^2$ (anti-squeezing) and $(\Delta {P_m})^2$ (squeezing), respectively. 
          }
    \label{fig_lossless}
\end{figure}

Third, so far as we obtain different bases, it is important to quantify the difference between them.
To do so, we compute the overlap matrix $\chi(U, V)$ of the basis sets $U$ and $V$ with matrix elements
\begin{equation}
    \chi_{ij}(U, V) =  \sum_k U_{ik} (V^H)_{kj}.
    \label{eq_overlap}
\end{equation}

Fourth, the purity of quantum Gaussian state with covariance matrix $\sigma$ is~\cite{Serafini_2004}
\begin{equation}
    \mathcal{P}(\sigma) = \dfrac{1}{\sqrt{\mathrm{det}(\sigma) }}.
    \label{eq_purity}
\end{equation}

Fifth, note that in the lossless case ($\alpha = 0$), the Mercer-Wolf, Williamson-Euler, and MSq modes coincide with the Schmidt modes and can be used as an alternative way to perform the BMr.

\section{Numerical Results}
\label{sec_results}

In this section, we present numerical studies of the influence of internal losses on spectral and squeezing properties of PDC.

To obtain an intuitive understanding of the influence of losses on PDC generation, we need a nonlinear medium, which satisfies the following conditions:
First, to demonstrate that our approach and equations are universal and suitable for arbitrary broadband pulsed high-gain PDC, we consider a short pump pulse and long nonlinear medium, which provides a temporal chirp both for the pump and PDC field.
Secondly, to show the difference between lossless, frequency-uniform, and frequency-dependent losses, we need a medium, that supports all three configurations with the same refractive indices.
The use of any concrete physical system and parameters does not allow us to make direct comparisons between three mentioned cases and would shift the focus of the investigation.

Therefore, in this paper, we use an artificial model of a dispersive material  with manually included losses.
We consider three cases of broadband high-gain PDC: lossless, frequency-uniform, and frequency-dependent losses.

For the lossless case, we consider the type-I ($e\rightarrow oo$) frequency-degenerate PDC in a nonlinear medium of $1$~cm length.
The refractive indices for $e$- and $o$-waves are given in Appendix~\ref{appendix_ref_index}.
We consider a Gaussian pulsed pump with a central wavelength of $800$~nm and a pulse duration of $50$~fs.
For these parameters, the low-gain two-photon amplitude is presented in Fig.~\ref{fig_lossless}(a).

\begin{figure*}
    \includegraphics[width=0.99\linewidth]{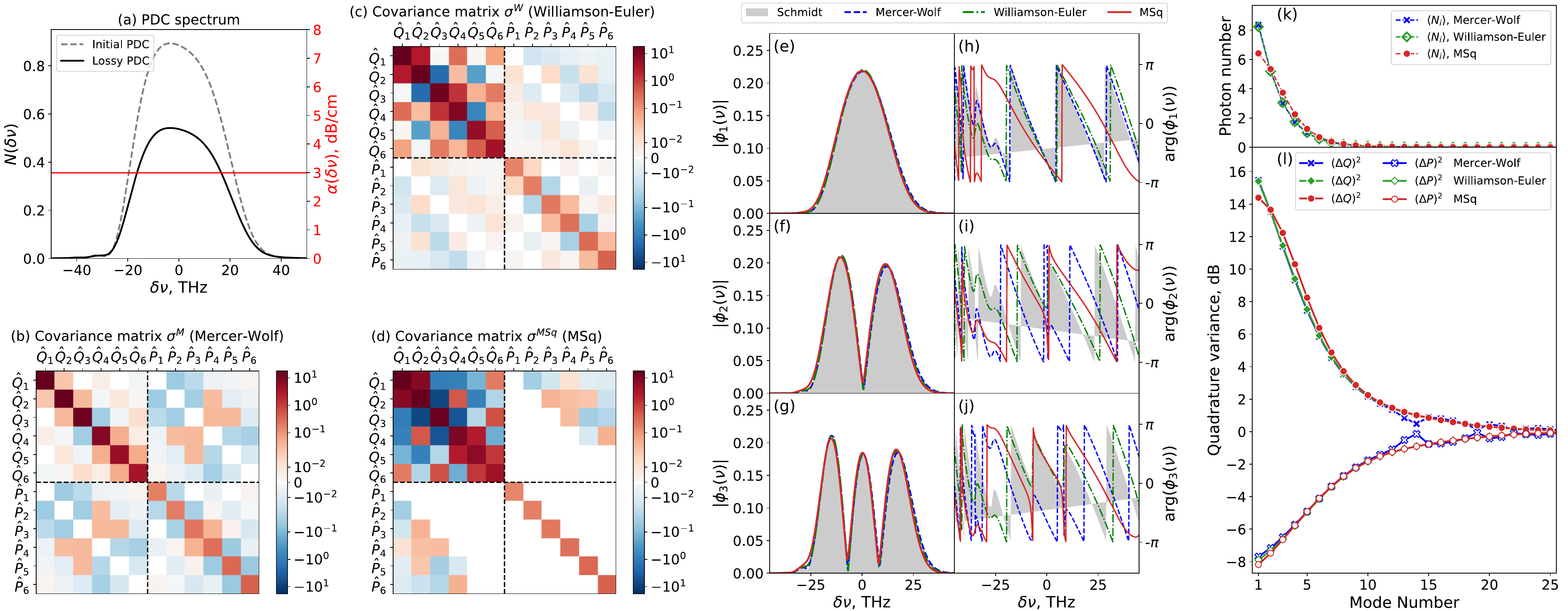 } 
        \caption{ 
        Numerical results for PDC with constant losses of $\alpha=3$~dB/cm (Example~1). The parameters of the model are given in the main text.  
        (a) PDC spectrum with losses (black solid line) compared to the initial lossless PDC (gray dashed line). The red line corresponds to the used extinction coefficient $\alpha$.
        (b, c, d) Covariance matrices of generated light  in the Mercer-Wolf, Williamson-Euler, and MSq bases, respectively.
        (e, f, g) Absolute values of the first (e), second (f), and third (g) PDC modes in the Mercer-Wolf (blue dashed), Williamson-Euler (green dot-dashed), and MSq (solid red) bases compared to the Schmidt modes of lossless PDC (grey shaded region).
        (h, i, j) Phases of the first (h), second (i), and third (j) PDC modes in the Mercer-Wolf (blue dashed), Williamson-Euler (green dot-dashed), and MSq (solid red) bases compared to the Schmidt modes of lossless PDC (grey shaded region).
        (k) Photon number distribution per mode in different bases.
        (l) Quadrature variances for the first 25 modes for the Mercer-Wolf (blue crosses), Williamson-Euler (green diamonds), and MSq (red circles) bases. Filled and empty markers correspond to $(\Delta {Q_m})^2$ (anti-squeezing) and $(\Delta {P_m})^2$ (squeezing), respectively.
         }
    \label{fig_uniform}
\end{figure*}

To increase the squeezing, we move to the high-gain regime, where the coupling constant $\Gamma$ is chosen such that the number of photons in the first Schmidt mode is equal to $N_1 = 14$, and the quadrature variance is $(\Delta P_1)^2=-17.6$~dB.
We solve Eqs.~\eqref{eq_diff_Efunc} and \eqref{eq_diff_Ffunc} numerically on a frequency grid of the size $N=511$ using the fourth-order Runge-Kutta method.
Using numerical Bogoliubov transfer matrices $E$ and $F$, we compute the covariance matrix in a monochromatic basis; this matrix is shown in Fig.~\ref{fig_lossless}(b).

Providing the BMr (Eq.~\eqref{eq_bloch_messiah_reduction}), we obtain the Schmidt-mode basis $U_{\mathcal{S}}$ with the number of occupied Schmidt modes $K_{\mathcal{S}} = 3.68$ for the above parameters.
The covariance matrix in the Schmidt-mode basis (Eqs.~\eqref{eq_covariance_matrix} and \eqref{eq_symplectic_transform}) is shown in Fig.~\ref{fig_lossless}(c).
It can be seen that this matrix is diagonal: There are no correlations between different Schmidt modes. 
The diagonal elements of the covariance matrix give us information about both anti-squeezing (quadrature variance $(\Delta Q)^2$) and squeezing (quadrature variance $(\Delta P)^2$) in different Schmidt modes (Fig.~\ref{fig_lossless}(d)).
Note that the state in each Schmidt mode is pure ($\Delta P_i \Delta Q_i  = 1$ for each mode $i$). 

\subsection{Example~1: constant losses}
\label{subsec_results_const}

In this example, we consider equal and constant losses of a magnitude $\alpha=3$~dB/cm for all frequencies of PDC.
All other parameters remain the same as for the lossless PDC.

For the given losses, we solve the master equations Eqs.~\eqref{eq_correlation_equation_1} and \eqref{eq_correlation_equation_2} numerically for the same frequency grid as for the lossless PDC and obtain the output correlation matrices $\braket{\hat{\mathbf{a}}^\dagger\hat{\mathbf{a}}}$ and $\braket{\hat{\mathbf{a}}\hat{\mathbf{a}}}$. 
For the numerical realization see Ref.~\cite{kopylov_2025_code}. 
The diagonal elements of $\braket{\hat{\mathbf{a}}^\dagger\hat{\mathbf{a}}}$ define the mean photon number distribution over frequency (spectrum), which is presented 
in Fig.~\ref{fig_uniform}(a) for both the lossless and lossy cases.
As expected, the intensity of PDC light generated in the lossy medium is lower compared to the lossless case.

\begin{figure*}
    \includegraphics[width=0.99\linewidth]{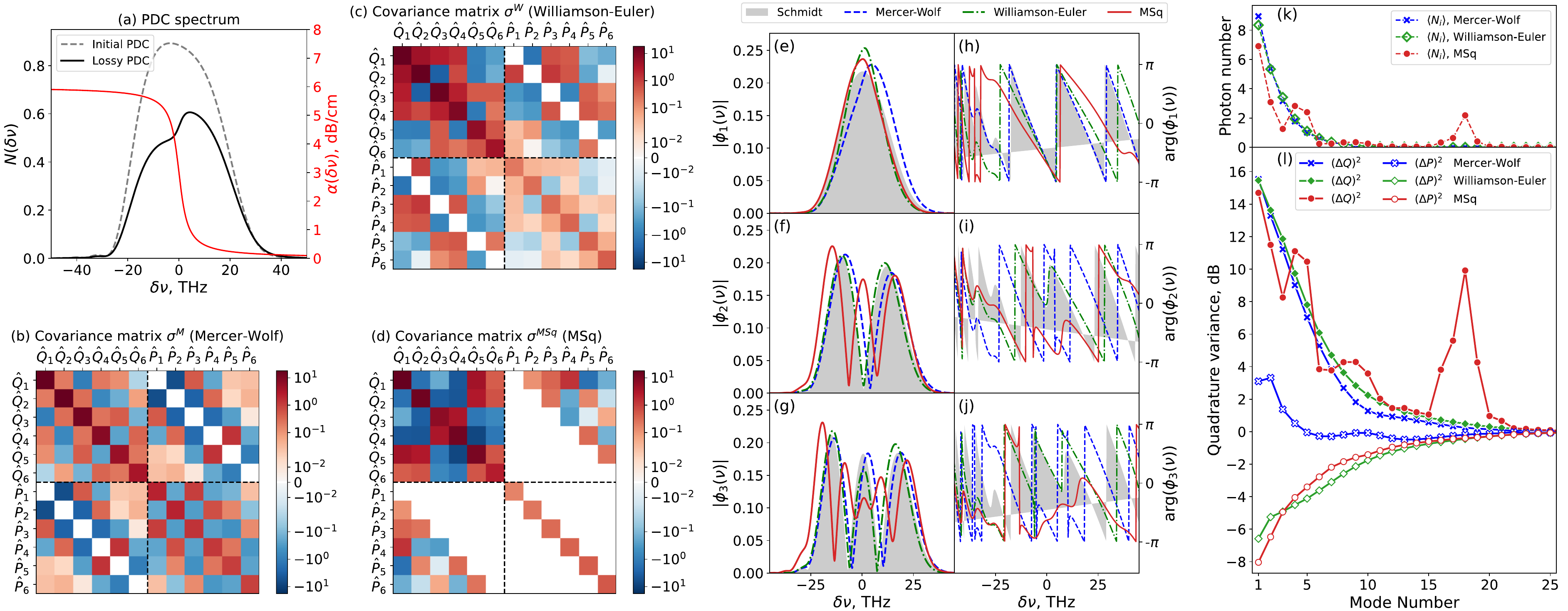} 
        \caption{
        Numerical results for PDC with frequency-dependent losses  of $\alpha(\omega)$ (Example~2). The parameters of the model are given in the main text.  
        (a) PDC spectrum with losses (black solid line) compared to the initial lossless PDC (gray dashed line). The red line shows the profile of the extinction coefficient $\alpha(\omega)$.
        (b, c, d) Covariance matrices of generated light in the Mercer-Wolf, Williamson-Euler, and MSq bases, respectively.
        (e, f, g) Absolute values of the first (e), second (f), and third (g) PDC modes in the Mercer-Wolf (blue dashed), Williamson-Euler (green dot-dashed), and MSq  (solid red) bases compared to the Schmidt modes of lossless PDC (grey shaded region).
        (h, i, j) Phases of the first (h), second (i), and third (j) PDC modes in the Mercer-Wolf (blue dashed), Williamson-Euler (green dot-dashed), and MSq (solid red) bases compared to the Schmidt modes of lossless  PDC (grey shaded region).
        (k) Photon number distribution per mode  in different bases.
        (l) Quadrature variances for the first 25 modes for the Mercer-Wolf (blue crosses), Williamson-Euler (green diamonds), and MSq (red circles) bases. Filled and empty markers correspond to $(\Delta {Q_m})^2$ (anti-squeezing) and $(\Delta {P_m})^2$ (squeezing), respectively.
        }
    \label{fig_nonuniform}
\end{figure*}

Having correlation matrices $\braket{\hat{\mathbf{a}}^\dagger\hat{\mathbf{a}}}$ and $\braket{\hat{\mathbf{a}}\hat{\mathbf{a}}}$, one can compute broadband bases mentioned above, namely the basis of Mercer-Wolf modes $U_M$, the Williamson-Euler basis $U_W$, and the MSq basis $U_{MSq}$, and construct the corresponding covariance matrices.
Fig.~\ref{fig_uniform}(b,c,d) shows the correlation matrices for the first six modes in the $U_M$, $U_W$, and $U_{MSq}$ bases, respectively.
It can be seen that, unlike for the lossless case, for lossy PDC the covariance matrix is not diagonal across all bases, which means that correlations between different modes are present.
The ellipses for each mode are aligned and $\braket{    \hat{P}_n \hat{Q}_n + \hat{Q}_n \hat{P}_n } = 0$. 
However, note that for $n \neq m$ the quadrature correlations $\braket{ \hat{P}_n \hat{Q}_m + \hat{Q}_m \hat{P}_n }$ are not necessarily zero.   
Fig.~\ref{fig_uniform}(e,f,g) and Fig.~\ref{fig_uniform}(h,i,j) present the absolute values and phases of the first three modes for different bases. 
Although the absolute values of the modes are very close to each other (and to the absolute values of the initial lossless Schmidt modes), their phases differ significantly and determine the main difference between the three bases.

As expected, losses decrease squeezing: Compared to the lossless variance $(\Delta P_1)^2=-17.6$~dB, the best squeezing of the lossy configuration is $(\Delta P_1)^2=-8.2$~dB and corresponds to the MSq basis. 
However, as shown in Fig.~\ref{fig_uniform}(k,l), the Mercer-Wolf and Williamson-Euler bases have similar photon number distributions and give squeezing values very close to the maximal squeezing.

\subsection{Example~2: frequency-dependent losses}
\label{subsec_results_const_fr_dep}
In this example, we consider frequency-dependent losses  $\alpha(\omega)$ with the profile shown by the red line in Fig.~\ref{fig_nonuniform}(a).
All other parameters remain the same as for the lossless PDC.
Performing the same steps as in the previous subsection, we present the PDC spectrum in Fig.~\ref{fig_nonuniform}(a) and the correlation matrices in the different bases (Mercer-Wolf, Williamson-Euler, and MSq) in Figs.~\ref{fig_nonuniform}(b,c,d), respectively.

It can be seen that the frequency-dependent losses change both the shape of the PDC spectrum and the profiles of the broadband modes. 
Fig.~\ref{fig_nonuniform}(e,f,g) presents the first three modes for the different basis sets: It can be seen that the Mercer-Wolf and Williamson-Euler modes are quite close to the initial lossless Schmidt modes, while the shape of the MSq-modes differs significantly.

Such a difference in the mode shapes leads to differences in the photon number distributions (Fig.~\ref{fig_uniform}(k)) and  squeezing (Fig.~\ref{fig_uniform}(l)) between bases. 
Indeed, as in the previous example, the best squeezing is found for the first MSq mode and constitutes  $(\Delta P_1)^2=-8.0$~dB. 
However, the best squeezing in the two other bases differs significantly: While the first  Williamson-Euler mode is characterized by $(\Delta P_1)^2=-6.6$~dB, the first Mercer-Wolf mode does not have squeezing at all, some squeezing appears only for high-order modes. 
This means that for the studied example, the Mercer-Wolf basis is not valid and the Williamson-Euler basis is not optimal for squeezing measurements.

\subsection{Discussion}

Here, we provide an overview of the properties of the discussed basis sets for constant and frequency-dependent losses. 

\begin{figure}
      \includegraphics[width=1.\linewidth]{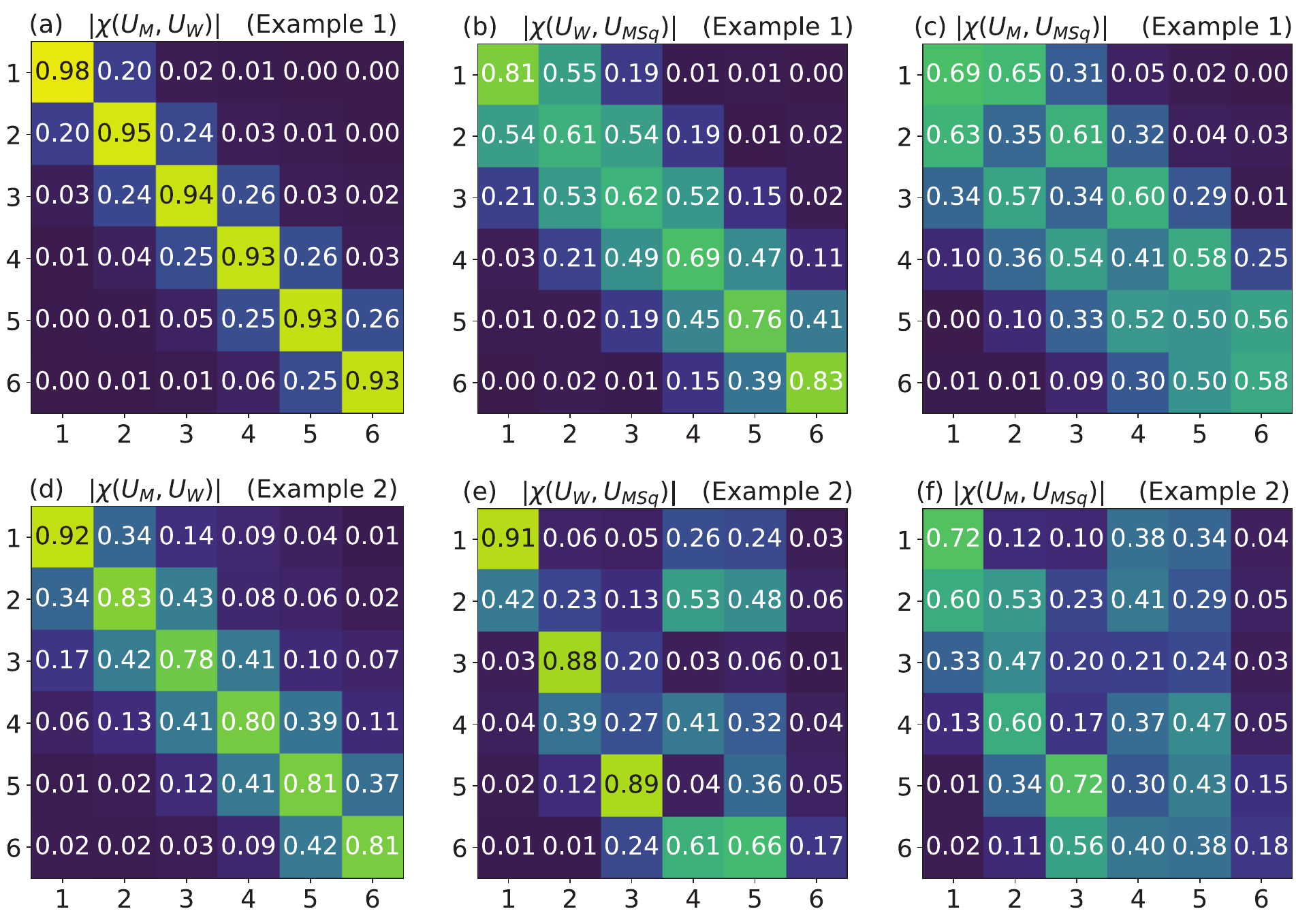 } 
    \caption{ Overlap matrices $\chi$ given by Eq.~\eqref{eq_overlap} for different basis sets.  (a,b,c) Example~1 and (d,e,f) Example~2.
       }
    \label{fig_overlaps}
\end{figure}

Table~\ref{table_summary} provides a summary of the main properties of the discussed bases. 
In the Mercer-Wolf basis, the output state has the lowest number of occupied modes; however, these modes do not provide maximal squeezing. 
The latter is also true for the Williamson-Euler basis. 
In turn, the first mode of the developed MSq-mode basis is characterized by the best squeezing in the lossy system. 
The overlap coefficients between studied bases are presented in Fig.~\ref{fig_overlaps}.
It can be seen that the Mercer-Wolf and Williamson-Euler modes are closer to each other when compared to the MSq modes.

The mismatch in mode-overlaps is more prominent for frequency-dependent losses: PDC photons for lower and higher frequencies reveal different losses, which leads to the skew of the PDC spectrum (see Fig.~\ref{fig_nonuniform}(a)).
The low mode-overlaps imply a significant difference in squeezing for the different bases. 

The measurement in a single broadband mode via the local oscillator can be interpreted as an operation on a single-mode subsystem from the whole generated multimode state, i.e., a reduced single-mode covariance matrix $\sigma^{(1)}$ describes all the measurable quantities.
Having the local oscillator with a shape of the first MSq-mode allows us to measure the maximal squeezing in the system. 
Compared to the single-mode subsystems built from first-modes from other bases, the single-mode subsystem from the first MSq-mode provides the maximal purity $\mathcal{P}_1 = \mathcal{P}(\sigma^{(1)})$.

The importance of the MSq-basis becomes more prominent, when we are able to measure in several modes, e.g. with demultiplexing.
Having two- and three-measurement modes, the measured state can be described by reduced covariance matrices $\sigma^{(12)}$ and  $\sigma^{(123)}$, respectively. 
The shape of the measurement modes determines an explicit form of reduced covariance matrices.  
Table~\ref{table_summary} contains the purities of reduced covariance matrices for the two-mode subsystem $\mathcal{P}_{12} = \mathcal{P}(\sigma^{(12)})$ and for the three-mode subsystem  $\mathcal{P}_{123} = \mathcal{P}(\sigma^{(123)})$ for the Mercer-Wolf, the Williamson-Euler, and the MSq bases.
For a fixed basis, an increase of the number of modes in the reduced state leads to a decrease of the purity: $\mathcal{P}_{1} > \mathcal{P}_{12} > \mathcal{P}_{123}$.
This means that each new mode adds additional noise, which reduces the total purity of subsystem.
Note that the purity of the full state (when all modes are taken into account) does not depend on the chosen basis. 

The MSq-basis shows better results for purities compared to Mercer-Wolf and Williamson-Euler modes. 
The explanation for such behavior can be found in Fig.~\ref{fig_nonuniform}(k,l).
For modes from 15 until 20 in the MSq-basis, there is a peak in the number of photons and $Q$-quadrature. 
It illustrates that the recursive procedure, which is used for the construction of the MSq-basis, shifts a significant amount of noise to the higher modes. 
In the Mercer-Wolf and the Williamson-Euler bases, such a peak is missing, therefore the same amount of noise is distributed over all the modes.
As a result, the MSq-basis provides less noise in lower modes and higher purities $\mathcal{P}_1$, $\mathcal{P}_{12}$ and $\mathcal{P}_{123}$, compared to other considered bases.

\begin{table}[h!]
\fontsize{8pt}{8pt}\selectfont
\begin{tabular}{c|cccc}
 & Schmidt & Mer.-W.   & W.-Eul.   & MSq   \\
\hline
Unitary & $U_{S}$ & $U_{M}$  & $U_{W}$  &$U_{Sq}$\\
Diagonal $\sigma$ &Yes &No  &No  &No\\
Diagonal $\braket{\hat{\mathbf{A}}^\dagger\hat{\mathbf{A}}}$ &Yes &  Yes   &No  &No\\
Minimal $K$    &Yes &  Yes   &No  &No\\
Max. squeezing &Yes &No  &No  &  Yes \\
\hline
\hline
\multicolumn{5}{c}{Lossless PDC} \\
\hline
$K$               & \multicolumn{4}{c}{ 3.68}   \\
$(\Delta P)^2_{min}$, dB  & \multicolumn{4}{c}{-17.6}   \\
$ \mathcal{P}$ & \multicolumn{4}{c}{1}       \\
\hline
\hline
\multicolumn{5}{c}{Example~1: $\alpha=3$~dB/cm} \\
\hline
$K$    & $-$ & 3.89   & 3.94   & 4.70 \\
$(\Delta P)^2_{min}$, dB & $-$ & -7.7   & -7.9   & -8.2 \\
$ \mathcal{P}_1$      & $-$ & 0.41 & 0.42 & 0.49 \\
$ \mathcal{P}_{12}$   & $-$ & 0.20 & 0.21 & 0.25 \\
$ \mathcal{P}_{123}$  & $-$ & 0.11 & 0.12 & 0.14 \\
\hline  
\hline
\multicolumn{5}{c}{Example~2: $\alpha=\alpha(\omega)$~dB/cm} \\
\hline
$K$                       & $-$ & 3.83   & 4.10   & 6.05 \\
$(\Delta P)^2_{min}$, dB  & $-$ & -0.5$^*$    & -6.6   & -8.0 \\
$ \mathcal{P}_1$      & $-$ & 0.12 & 0.36 & 0.46 \\
$ \mathcal{P}_{12}$   & $-$ & 0.08 & 0.18 & 0.26 \\
$ \mathcal{P}_{123}$  & $-$ & 0.05 & 0.10 & 0.20 \\
\hline
 \end{tabular} \\
 \footnotesize{*-for the 13-th mode}
\caption{ Comparison of different bases for lossy PDC and the Schmidt bases of lossless PDC. 
Here, $(\Delta P)^2_{min}$ is the minimal quadrature variance, $K$ is the number of modes Eq.~\eqref{eq_number_of_modes}.
$ \mathcal{P}_1$, $ \mathcal{P}_{12}$, and $ \mathcal{P}_{13}$ are purities of subsystems consisting of only the first mode, the first two-modes, and the first three-modes from each basis, respectively.      
} 
\label{table_summary}
\end{table}

\section{Conclusions}
\label{sec_conclusions}

In this work, we examine the properties of multimode squeezed light in a lossy medium. 
Having the spatial Heisenberg equation and losses introduced via a discrete set of virtual beamsplitters, we show that the solution to the output annihilation operators has a form of partial Bogoliubov transformation.
In the limit of continuous spatially-uniform losses, we consider the spatial Langevin equation for the monochromatic operators, which is valid for any Markovian environment and frequency-dependent losses. 
For Gaussian states, we present the master equations for the second-order correlation matrices.
The spatial-ordering effects are intrinsic for the obtained Langevin and master equations; therefore, our approach is valid for broadband pump pulses and both for low- and high-gain PDC.

For lossy PDC, we present the analysis of its mode structure.
Unlike the lossless case, where the Bloch-Messiah reduction results in a unique basis of Schmidt modes, internal losses lead to quadrature correlations for any broadband basis. 
Considering the task of finding a broadband basis for maximizing squeezing, we investigate the Mercer-Wolf and Williamson-Euler modes and show that they are not optimal. 
In contrast, we present a basis of broadband modes, which allows us to obtain the maximal squeezing in the considered system. This basis is unique for any lossy system, and we propose an algorithm to find it.

For multimode PDC generated in a lossy medium, any broadband basis provides correlations between quadratures of different modes, i.e., the covariance matrix can not be diagonalized. 
Consequently, there is no unique broadband basis of Schmidt modes as for the lossless case. 
The proper broadband basis should be chosen with respect to the considered scientific problem or desired applications.
The choice of a basis is not limited to the Mercer-Wolf, Williamson-Euler, or MSq basis, e.g., the broadband modes can also be considered from the diagonalization of the $\braket{\hat{\mathbf{a}}\hat{\mathbf{a}}}$ matrix using the Takagi factorization or by using optimization techniques in order to maximize or minimize desired physical quantities.

The methods and approaches presented in this paper are not limited to type-I PDC: All the procedures can be performed for other PDC types by modifying the vectors of annihilation and creation operators in accordance with their polarization.
Although as an example we consider the rather artificial case of frequency-dependent PDC losses, all the presented equations and expansions are also valid for pulsed single-pass four-wave mixing in resonant media, where the absorption spectrum of the media, which may have a complex profile, plays a crucial role~\cite{Swaim_2017}. 

\begin{acknowledgments}
We would like to thank  M. Stefszky, H. Herrmann, L. Padberg, S. Barkhofen and Prof. C. Silberhorn for fruitful discussions, which helped to significantly improve the quality of the manuscript.
This work is supported by the ”Photonic Quantum Computing” (PhoQC) project, which is funded by the Ministry for Culture and Science of the State of North-Rhine Westphalia.
\end{acknowledgments}

\bibliographystyle{quantum}
\bibliography{lit_pdc_losses.bib}

\onecolumn
\appendix

\section{Solution of the Heisenberg equation for lossless PDC}
\label{appendix_lossless_pdc}

For a quantum description of a single-pass PDC generation in a finite nonlinear crystal we use the spatial generator formalism~\cite{Huttner_1990,Horoshko_2022}.
Under this approach, the light inside the dispersive medium is presented in terms of discrete temporal modes, with the frequencies $\omega_m =  2\pi m/T $, where $m = 0, 1, 2 \dots$, and the discretization time $T$ is much longer than any relevant  time in the system.

The frequency positive part of the spatially single-mode electric field operator has the form
\begin{equation}
\hat{E}^+(z, t) = \sum_{n} \sqrt{\dfrac{\hbar \omega_n}{2 \varepsilon_0 c T n(\omega_n)}}  \hat{a}(z, \omega_n) e^{- i \omega_n t},
\label{ap_eq_field_quantized}
\end{equation} 
where $n(\omega)$ is a refractive index. The annihilation $\hat{a}(z, \omega_n)$ and creation $\hat{a}^\dagger(z, \omega_n)$ operators satisfy the bosonic commutation relation $[\hat{a}(z, \omega_n), \hat{a}^\dagger(z, \omega_m)]=\delta_{nm}$.

The spatial Heisenberg equation has the form~\cite{Huttner_1990,Horoshko_2022} 
\begin{equation}
i\hbar\dfrac{\partial \hat{a}(z, \omega_n)}{\partial z} = [\hat{a}(z, \omega_n), -\hat{G}(z)], 
\label{eq_ap_heisenberg}
\end{equation} 
where the momentum operator $\hat{G}(z)$ is a generator for spatial evolution.

For the type-I PDC generated by a pulsed pump of the form
\begin{eqnarray}
E^+_p(z, t) = \int d\omega \ S(\omega)e^{ -i\omega t  + i k_p(\omega)z},
\end{eqnarray}
the momentum operator is given by $\hat{G}(z)=\hat{G}_{l}(z)+\hat{G}_{nl}(z)$,  where
\begin{equation}
\hat{G}_{l}(z) =  \sum_{n}  \hbar k_n \hat{a}^\dagger(z, \omega_n) \hat{a}(z, \omega_n)  + h.c. 
\label{eq_ap_momentum_lin}
\end{equation} 
and
\begin{equation}
\hat{G}_{nl}(z) =   \dfrac{\hbar\Gamma}{2} \sum_{i,j} J_{ij}(z)  \hat{a}^\dagger(z, \omega_i) \hat{a}^\dagger(z, \omega_j)  + h.c.
\label{eq_ap_momentum_nonlin}
\end{equation} 
Here, the coupling matrix $J_{ij}(z) = S(\omega_i + \omega_j) e^{i k_p(\omega_i + \omega_j)z} $, 
$k_n \equiv k(\omega_n) = \frac{n(\omega_n)\omega_n}{c}$ is the wavevector of quantized modes in a crystal,
$S(\omega)$ is the pump spectrum  at $z=0$, and $k_p(\omega)=\frac{n_p(\omega)\omega}{c}$ is the pump wavevector.

Substituting Eqs.~\eqref{eq_ap_momentum_lin} and~\eqref{eq_ap_momentum_nonlin} in Eq.~\eqref{eq_ap_heisenberg}, the spatial Heisenberg equation for PDC monochromatic operators takes the form
\begin{equation}
\dfrac{d \hat{a}_i(z)}{dz}   = i k_i \hat{a}_i(z) + i\Gamma \sum_{j} J_{ij}(z) \hat{a}_j^\dagger(z).
\label{eq_lossless_Heisenberg}
\end{equation} 

The solution to the Heisenberg equation has the form of the Bogoliubov transformation: 
\begin{equation}
        \hat{a}_i(z) = \sum_j \Big( E_{ij}(z, z_0) \hat{a}_j(z_0) +  F_{ij}(z, z_0) \hat{a}^\dagger_j(z_0)  \Big),
        \label{eq_bogoliubov_full_app}
 \end{equation}
where $z_0$ is the initial position of the nonlinear medium.
The Bogoliubov transfer matrices satisfy the following equations
\begin{align}
    \dfrac{d E_{ij}(z, z_0)}{dz}  = i k_i E_{ij}(z) + i\Gamma \sum_{n} J_{in}(z) F^{*}_{nj}(z, z_0),
    \label{eq_diff_Efunc}  
    \\    
    \dfrac{d  F_{ij}(z, z_0)  }{dz}  = i k_i F_{ij}(z) + i\Gamma \sum_{n} J_{in}(z)  E^{*}_{nj}(z, z_0).
    \label{eq_diff_Ffunc}
\end{align}
The  initial conditions read $E_{ij}(z_0,z_0)=\delta_{ij}$ and $F_{ij}(z_0,z_0)=0$.

In order to solve these equations numerically, one should get rid of fast oscillating terms. 
This can be done by introducing the slowly-varying functions $\mathcal E_{ij}(z)$ and $\mathcal F_{ij}(z)$
\begin{align}
    E_{ij}(z) = \mathcal E_{ij}(z,z_0)e^{i k_i z},
    \\
    F_{ij}(z) = \mathcal F_{ij}(z,z_0) e^{i k_i z}.
\end{align}
Then, the slowly-varying equations read
\begin{align}
    \dfrac{d  \mathcal E_{ij}(z)  }{dz}   = i\Gamma \sum_{n} S(\omega_i + \omega_n)e^{i \Delta k_{in} z} \mathcal F^{*}_{nj}(z),   
    \label{ap_eq_slowly_eq1}
    \\    
    \dfrac{d   \mathcal F_{ij}(z)  }{dz}  = i\Gamma \sum_{n} S(\omega_i + \omega_n)e^{i \Delta k_{in} z}  \mathcal E^{*}_{nj}(z), 
    \label{ap_eq_slowly_eq2}
\end{align}
where $\Delta k_{ij} = k_p(\omega_i + \omega_j) - k_i - k_j$ is the wave-vector mismatch.
Note that the wave-vector mismatch $\Delta k_{ij}$  in the equations appears only for the slowly-varying functions and operators.
From numerical point of view, these slowly-varying equations can be solved in much more efficient way compared to original Eqs.~\eqref{eq_diff_Efunc} and \eqref{eq_diff_Ffunc}.

For given Bogoliubov transfer matrices, the field correlation functions for a vacuum initial state can be found as
\begin{align}
    \braket{\hat{a}^\dagger_i(z) \hat{a}_j(z)} &~= ~ \sum_n  ({F}_{in})^{*} {F}_{jn}, \\
    \braket{\hat{a}_i(z) \hat{a}_j(z)}  &~= ~ \sum_{n} {E}_{in} {F}_{jn}. 
\end{align}

\subsection{Approximate spatially-averaged solution}

One of the ways to solve Eqs.~\eqref{ap_eq_slowly_eq1} and \eqref{ap_eq_slowly_eq2} approximately is to replace the function $e^{i \Delta z}$ with its spatially-averaged version 
\begin{equation}
    e^{i \Delta z} \rightarrow \frac1{L}\int_0^L dz e^{i \Delta z} = \mathrm{sinc}\Big(\frac{\Delta L}{2}\Big)e^{i\frac{\Delta L}{2}}.
\end{equation}
In this case, the equations~\eqref{ap_eq_slowly_eq1} and \eqref{ap_eq_slowly_eq2} became autonomous and their solution can be easily found using theories with the gain-independent Schmidt modes e.g. in Refs.~\cite{Sharapova_2015,Pe_ina_2015,Sharapova_2018}.
This spatially-averaged solution can also be interpreted as the first-order Magnus expansion of the evolution operator~\cite{Horoshko_2019},
which gives only a qualitative solution for the studied system.
Under this approximation, the spatial-ordering effects are neglected and consequently the local losses can not be properly implemented. 
Therefore, we do not use this approximation in this paper and solve the equations numerically with $z$-dependent coupling matrix $J(z)$.

\section{Full Bogoliubov transformation and Bloch-Messiah reduction}
\label{appendix_bogoliubov_full}

The full Bogoliubov transformation is
\begin{equation}
\begin{bmatrix} 
    \mathbf{\hat{b}} \\
    \mathbf{\hat{b}}^{\dagger} 
 \end{bmatrix}
 =
 \begin{bmatrix}  
        E        &  F\\ 
        F^{\ast} &  E^{\ast} 
 \end{bmatrix}
\begin{bmatrix} 
    \mathbf{\hat{a}} \\
    \mathbf{\hat{a}}^{\dagger} 
 \end{bmatrix},
\end{equation}
where the matrices $E$ and $F$ are the square $N\times N$ matrices, while  $\mathbf{\hat{a}}$, $\mathbf{\hat{b}}$ are the vectors of the input and output annihilation operators, respectively.

The output operators satisfy the bosonic commutation relations that  results in the following relations for the matrices:
\begin{equation}
    E E^H - F F^H = \mathbf{1}_N, ~ E F^T = (E F^T)^T,
    \label{ap_1_commut}
\end{equation} 
where  $\mathbf{1}_N$ is the identity matrix.

The full Bogoliubov transformation is invertible, so it is possible to write the input commutation relations through the output operators that leads to two additional relations:
\begin{equation}
    E^H E  - (F^H F)^T = \mathbf{1}_N, ~ E^H F = (E^H F)^T.
    \label{ap_1_commut_inverse}
\end{equation}

In order to obtain the Bloch-Messiah reduction, we perform the left polar decomposition of the matrices $E$ and $F$:
\begin{equation}
      E = P_E V_{E}, ~ ~  F = P_F V_{F},
\end{equation}
where the matrices $V_{E}$ and $V_{F}$ are the unitary matrices. 
From the definition, the matrices $P_E = (EE^H)^{\frac12}$ and $P_F = (FF^H)^{\frac12}$ are the square positive definite matrices.
According to Eq.~\eqref{ap_1_commut}, matrices $EE^H$ and $FF^H $ commute;
therefore, matrices $P_F$ and $P_E$ are simultaneously diagonalizable with the eigendecomposition
\begin{equation}
P_E  = U \Lambda_E U^H, ~ ~   P_F = U \Lambda_F U^H,
\label{eq_ap_1_eigdec}
\end{equation}
where $\Lambda_E$ and $\Lambda_F$ are the real and non-negative matrices. Their  values are sorted in descending order and, according to Eq.~\eqref{ap_1_commut},  are connected as $\Lambda_E^2 = \Lambda_F^2 + \mathbf{1}_N$.

Therefore, the matrices $E$ and $F$ can be written in the form of simultaneous singular value decomposition
\begin{align}
        E = U \Lambda_E U^H V_{E} = U \Lambda_E W_E^H, \\
        F = U \Lambda_F U^H V_{F} = U \Lambda_F W_F^H.
\end{align}

Since the eigendecomposition~\eqref{eq_ap_1_eigdec} is not unique, there is a freedom in choosing matrix $U$ (the eigenvectors are defined up to an arbitrary phase).
However, the presence of the inverse transformation~\eqref{ap_1_commut_inverse} allows us to get rid of this phase-freedom by applying the so-called the `rotation condition' $D = (W_F^H W_E^*)^{\frac12}$~\cite{Cariolaro_2016,Cariolaro_2016_takagi}.

The resulting decomposition of the Bogoliubov transformation has the form of the Bloch-Messiah reduction~\cite{Braunstein_2005} 
\begin{equation}
        E =  \mathcal U \Lambda_E \mathcal W_E^H, ~ ~ ~ 
        F =  \mathcal U \Lambda_F  ( \mathcal W_E^H)^*,
\end{equation}
where $\mathcal U = U D$,  $\mathcal W_E = W_F^* D^*$ and $\mathcal W_F = W_F D$.

Note that the Bloch-Messiah reduction is valid only for the closed system (full Bogoliubov transformation), where the invertible transformation can be constructed.

\section{Number of occupied modes}
\label{appendix_prove_occ_modes}

In this section, we show that the number of occupied modes defined as
\begin{equation}
    K_F = \Big( \sum_i (n^F_i)^2 \Big)^{-1},
\end{equation}
where 
$n^F_i=\braket{\hat{F}_i^\dagger \hat{F}_i}/(\sum_i \braket{\hat{F}_i^\dagger \hat{F}_i})$ is minimal for modes having a diagonal correlation matrix $ \braket{\hat{F}_i^\dagger  \hat{F}_j }$.

To prove it, let us consider two sets of modes $\hat{A}$ and $\hat{B}$.
The first one provides the diagonal correlation matrix $ \braket{\hat{A}_i^\dagger  \hat{A}_j } = \lambda^2_i \delta_{ij} $, where $\lambda_i^2$ are real and non-negative values.
Modes $\hat{B}$ are connected with modes $\hat{A}$ by the unitary matrix $U$:
\begin{align}
        \hat{B}_n  = \sum_m U_{nm} \hat{A}_m.
\end{align}

The diagonal elements of correlation matrices in basis $\hat{B}$ can be found using the transformation~\eqref{eq_correlation_transformation}:
 \begin{equation}
    \braket{\hat{B}^\dagger_n \hat{B}_n } = \sum_m U^*_{nm} \lambda_m^2 U_{nm} =  \sum_m |U_{nm}|^2  \lambda_m^2.
\end{equation}

Then, the full number of photons reads
\begin{equation}
    \mathcal{N} = \sum_n \braket{\hat{A}^\dagger_n \hat{A}_n } = \sum_n \braket{\hat{B}^\dagger_n \hat{B}_n } = \sum_n \lambda_n^2.
\end{equation}

At the same time, since $ 0 \leq|U_{nm}|^2 \leq 1$, the following inequality holds
\begin{equation}
    \sum_n (n^{B}_n)^2 = \dfrac{1}{\mathcal{N}^2}\sum_n   \big( \sum_m |U_{nm}|^2  \lambda_m^2 \big)^2 \leq\dfrac{1}{\mathcal{N}^2}   \sum_n \big( \sum_m \lambda_m^2 \big)^2   = \sum_n (n^{A}_n)^2.
\end{equation}

Therefore,
\begin{equation}
    K_A \leq K_B.
\end{equation}

\section{Construction of a discrete solution}
\label{appendix_discrete_model}

\subsection{Transfer matrix method for the Bogoliubov transformation}

Let us consider the Bogoliubov transformation in the form
\begin{equation}
   \begin{bmatrix}
        \hat{\mathbf{a}}(z)      \\
        \hat{\mathbf{a}}^\dagger(z) 
  \end{bmatrix} 
= \mathcal{M}(z,z_0) 
   \begin{bmatrix}
        \hat{\mathbf{a}}(z_0)      \\
        \hat{\mathbf{a}}^\dagger(z_0) 
  \end{bmatrix}.
  \label{eq_BT_general}
\end{equation}
Here, $\hat{\mathbf{a}}(z) = [\hat{a}_1(z), \hat{a}_2(z), \dots,  \hat{a}_N(z)]^T$ and $\hat{\mathbf{a}}^\dagger(z) = [\hat{a}^\dagger_1(z), \hat{a}^\dagger_2(z), \dots,  \hat{a}^\dagger_N(z)]^T$ are the vectors consisting of the  annihilation and creation operators, respectively.
The matrix $\mathcal{M}(z,z_0) $ is the $2N\times 2N$ Bogoliubov transfer matrix of the full system and can be written as  
\begin{equation}
    \mathcal{M}(z,z_0) = 
   \begin{bmatrix}
        \mathcal{M}_e(z,z_0)     & \mathcal{M}_f(z,z_0)   \\
        \mathcal{M}_f^*(z,z_0)   & \mathcal{M}_e^*(z,z_0) \\
  \end{bmatrix} , 
\end{equation}
where the matrices $\mathcal{M}_e$ and $\mathcal{M}_f$ are $N\times N$ sub-matrices. 
 $ \mathcal{M}(z,z_0)$ matrices couple the annihilation operators at position $z$ with the creation and annihilation operators at position $z_0$.

If we introduce the annihilation and creation operators at position $z^\prime$,  where $z_0 < z^\prime < z$, and two intermediate Bogoliubov transformations $\mathcal{M}(z, z^\prime)$ and $\mathcal{M}(z^\prime, z_0)$, then the full transformation (from $z_0$ to $z$) is given by $\mathcal{M}(z,z_0)  = \mathcal{M}(z, z^\prime)  \mathcal{M}(z^\prime, z_0) $, where the sub-matrices read 
\begin{equation}
    \mathcal{M}_e(z,z_0) =      \mathcal{M}_e(z, z^\prime)  \mathcal{M}_e(z^\prime, z_0) 
                   +\mathcal{M}_f(z, z^\prime)  \mathcal{M}_f^*(z^\prime, z_0),  
\label{eq_app_multrule_1}
\end{equation}
\begin{equation}
    \mathcal{M}_f(z,z_0) =      \mathcal{M}_e(z, z^\prime)  \mathcal{M}_f(z^\prime, z_0)  
                     +\mathcal{M}_f(z, z^\prime)  \mathcal{M}_e^*(z^\prime, z_0).
\label{eq_app_multrule_2}
\end{equation}

\subsection{ Discrete model: step by step }
\label{appendix_discrete_model__}

 \begin{figure}[h!]
    \centering
      \includegraphics[width=0.5\linewidth]{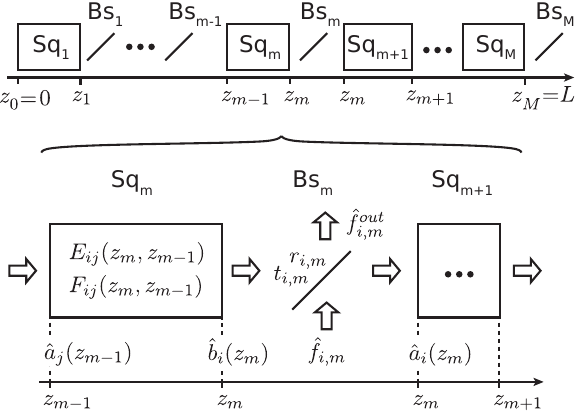}\\
    \caption{ Detailed scheme of the discrete model. }
    \label{fig_scheme_app_detailed}
\end{figure}

As was mentioned in Section~\ref{sec_pdc_descrete_losses}, a lossy medium with the extinction coefficient $\alpha(\omega)$ is modeled as an $M$-segmented nonlinear medium, where each segment of length $\Delta z$ consists of an ideal lossless squeezer, and beamsplitters are placed between such lossless segments (Fig.~\ref{fig_scheme_app_detailed}).

In terms of the Bogoliubov transformation, this system consists of the input operators $\hat{\mathbf{A}}_{in} = (\hat{\mathbf{a}}, \hat{\mathbf{f}}_1, \hat{\mathbf{f}}_2, \dots, \hat{\mathbf{f}}_M)^T$, and the output operators $\hat{\mathbf{A}} _{out}= (\hat{\mathbf{a}}(L), \hat{\mathbf{f}}^{out}_1, \hat{\mathbf{f}}^{out}_2, \dots, \hat{\mathbf{f}}^{out}_M)^T$.
The input and output operators are coupled with the full Bogoliubov transformation 
\begin{equation}
\begin{bmatrix} 
     \hat{\mathbf{A}}_{out} \\
     \hat{\mathbf{A}}_{out}^{\dagger} 
 \end{bmatrix}
 =
 \begin{bmatrix}  
        \mathcal{M}_e        &   \mathcal{M}_f \\ 
        \mathcal{M}_f^{\ast} &  \mathcal{M}_e^{\ast} 
 \end{bmatrix}
\begin{bmatrix} 
     \hat{\mathbf{A}}_{in} \\
     \hat{\mathbf{A}}_{in}^{\dagger} 
 \end{bmatrix}.
\end{equation}

However, as mentioned in the main text, we do not need to calculate the full matrices $\mathcal{M}_e$ and $\mathcal{M}_f$, rather only the part of these matrices that corresponds to the output operators $\hat{\mathbf{a}}(L)$.
So, the partial Bogoliubov transformation has the from 
\begin{equation}
\begin{bmatrix} 
    \hat{\mathbf{a}}(L) \\
    \hat{\mathbf{a}}^{\dagger}(L) 
 \end{bmatrix}
 =
 \begin{bmatrix}  
        \tilde{\mathbf{E}}        &   \tilde{\mathbf{F}} \\ 
        \tilde{\mathbf{F}}^{\ast} &  \tilde{\mathbf{E}}^{\ast} 
 \end{bmatrix}
\begin{bmatrix} 
    \hat{\mathbf{A}}_{in} \\
    \hat{\mathbf{A}}_{in}^{\dagger} 
 \end{bmatrix},
\end{equation}
where $ \tilde{\mathbf{E}} = (  \tilde E^{0} \:  \tilde E^{1} \:  \tilde E^{2} \dots \tilde E^{M})$ and  $\tilde{\mathbf{F}} = (  \tilde F^{0} \:  \tilde F^{1} \:  \tilde F^{2} \dots \tilde F^{M})$ are the partial Bogoliubov transformation matrices. 

The matrices $\tilde E^{m}$ and $\tilde F^{m}$ can be found in the following way: 

1. The $m$-th ideal squeezer with the initial and final coordinates  $z_{m-1}$ and $z_{m}$, respectively, is described by the Bogoliubov transfer matrices $E_{m} = E(z_{m}, z_{m-1})$ and $F_{m} = F(z_{m}, z_{m-1})$ that are the solutions to Eqs.~\eqref{eq_diff_Efunc} and~\eqref{eq_diff_Ffunc} with the initial conditions  $E(z_{m-1}, z_{m-1}) = \mathbf{1}_N$ and $F(z_{m-1}, z_{m-1}) = 0$.
The Bogoliubov transformation is then
 \begin{equation}
\begin{bmatrix} 
    \mathbf{b}(z_{m})         \\
    \mathbf{b}^{\dagger}(z_{m}) 
 \end{bmatrix}
 =
 \begin{bmatrix}  
        E_{m}        &   F_{m} \\ 
        F^{\ast}_{m} &  E^{\ast}_{m} 
 \end{bmatrix}
\begin{bmatrix} 
    \hat{\mathbf{a}}(z_{m-1}) \\
    \hat{\mathbf{a}}^{\dagger}(z_{m-1}) 
 \end{bmatrix}.
     \label{eq_ap_BTSQ}
\end{equation}

2. The beamsplitter matrices are given by
\begin{align}
    T_m &= \mathrm{diag}(t_{1,m}, t_{2,m},\cdots, t_{N,m}), \\    
    R_m &= \mathrm{diag}(r_{1,m}, r_{2,m},\cdots, r_{N,m}),
\end{align}
where $t_{i,m}  \equiv t_m(\omega_i) = e^{-\alpha_m(\omega_i)\Delta z/2}$ and $r_m(\omega_i) = \sqrt{1 - t^2_m(\omega_i)}$ are the transmission and reflection coefficients, respectively.
The Bogoliubov transformation for each beamsplitter reads
 \begin{equation}
\begin{bmatrix} 
    \hat{\mathbf{a}}(z_{m})         \\
    \hat{\mathbf{f}}^{out}_m   \\
 \end{bmatrix}
 =
 \begin{bmatrix}  
        T_{m}        &   R_{m} \\ 
        -R_{m}       &   T_{m} \\ 
 \end{bmatrix}
\begin{bmatrix} 
    \mathbf{b}(z_{m})         \\
    \hat{\mathbf{f}}_m   \\
 \end{bmatrix}.
\label{eq_ap_BTBS}
\end{equation}

For a single segment consisting of one squeezer and one beamsplitter, the multiplication of the Bogoliubov transformations
~\eqref{eq_ap_BTSQ} and~\eqref{eq_ap_BTBS} leads to the following expression for the annihilation operators
\begin{equation}
    \hat{a}_i(z_{m}) = t_{i,m} \Big( \sum_{n} E_{in}(z_{m}, z_{m-1})\hat{a}_n(z_{m-1}) \\ + F_{in}(z_{m}, z_{m-1})\hat{a}^\dagger_n(z_{m-1}) \Big) + r_{i,m}  \hat{f}_{i,m}.
    \label{eq_app_segment_solution}
\end{equation}

By multiplying all the segments with the rules~\eqref{eq_app_multrule_1} and~\eqref{eq_app_multrule_2}, we find the final matrices $\tilde E^{m}$ and $\tilde F^{m}$, connecting the operators at the beginning and the end of the crystal.
In the most compact way, these procedure can be represented using the following inverse recursive expression from $m=M$ to $m=0$ 
\begin{align}
    Y^{e}_{m} &= Y^{e}_{m+1} T_{m}  E_{m} +  Y^{f}_{m+1}  (T_{m}  F_{m})^*, \\
    Y^{f}_{m} &= Y^{e}_{m+1} T_{m}  F_{m} + Y^{f}_{m+1} (T_{m}  E_{m})^*, \\
    & \text{with }  Y^{e}_{M+1}  = \mathbf{1}_N  ~ ~ \text{and} ~ ~  Y^{f}_{M+1}  = 0.
    \label{eq_recursive}
\end{align}
The partial Bogoliubov transfer matrices from Eq.~\eqref{eq_partial_bogoliubov} are 
\begin{align}
\tilde E^{m} = Y^{e}_{m+1} R_{m},& ~ ~ ~
\tilde F^{m} = Y^{f}_{m+1} (R_{m})^*,\\
\tilde E^{0} = Y^{e}_{0},&  ~ ~ ~
\tilde F^{0} = Y^{f}_{0}.
\end{align}

The annihilation operators at the output of the crystal satisfy the bosonic commutation relations $[\hat{a}_i(L),\hat{a}^\dagger_j(L)]=\delta_{ij}$ and $[\hat{a}_i(L),\hat{a}_j(L)]=0$ and lead to additional relations for the matrices:
\begin{align}
     \sum_{m=0}^M  \Big[  \tilde{E}^{m}  (\tilde{E}^{m} )^{H} - \tilde{F}^{m} (\tilde{F}^{m} )^{H} \Big] =&  \ \delta_{ij}, \\
  \sum_{m=0}^M  \Big[ \tilde{E}^{m} (\tilde{F}^{m} )^T - \tilde{F}^{m}   (\tilde{E}^{m} )^T \Big]  =& \  0,
\end{align}
respectively.

Using the partial Bogoliubov transformation, one can calculate various correlation matrices.
For example, for the initial vacuum state and vacuum environment they read
 \begin{align}
      \braket{\hat{a}^\dagger_i(z) \hat{a}_j(z)} & \ = \sum_n  (\tilde{F}^{0}_{in})^{*} \tilde{F}^{0}_{jn}  + \sum_{m=1}^M  \sum_{n}  (\tilde{F}^{m}_{in})^{*}  \tilde{F}^{m}_{jn} ,
      \label{eq_ap_first_corr}
    \\
    \braket{\hat{a}_i(z) \hat{a}_j(z)}  & \ = \sum_{n} \tilde{E}^{0}_{in} \tilde{F}^{0}_{jn}  + \sum_{m=1}^M \sum_{n} \tilde{E}^{m}_{in} \tilde{F}^{m}_{jn}. 
      \label{eq_ap_second_corr}
\end{align}

It should be noted that the first sum in Eqs.~\eqref{eq_ap_first_corr} and~\eqref{eq_ap_second_corr} corresponds to the contribution of the initial vacuum state $\hat{\mathbf{a}}$, while the second sum appears due to the environment vacua $\hat{\mathbf{f}}_i$.

\section{Partial Bogoliubov transformation and its decomposition}
\label{appendix_bogoliubov_part}

Let us consider the partial Bogoliubov transformation in the form
\begin{equation}
\begin{bmatrix} 
    \mathbf{\hat{b}} \\
    \mathbf{\hat{b}}^{\dagger} 
 \end{bmatrix}
 =
 \begin{bmatrix}  
        \tilde E        &   \tilde F \\ 
        \tilde F^{\ast} &  \tilde E^{\ast} 
 \end{bmatrix}
\begin{bmatrix} 
    \mathbf{\hat{c}} \\
    \mathbf{\hat{c}}^{\dagger} 
 \end{bmatrix}.
\end{equation}
Here, the matrices $\tilde{E}$ and $\tilde{F}$ are $N\times N(M+1)$ matrices, the vector $\mathbf{\hat{c}}$ has the size of $N(M+1)$, while the vector $\mathbf{\hat{b}}$ has the size of $N$.

The output operators $\mathbf{\hat{b}}$ satisfy the bosonic commutation relations, so 
\begin{equation}
    \tilde{E} \tilde{E}^H - \tilde{F} \tilde{F}^H = \mathbf{1}_N, ~ \tilde{E} \tilde{F}^T = (\tilde{E} \tilde{F}^T)^T.
    \label{commut_part}
\end{equation} 

In order to perform a decomposition for the partial Bogoliubov transformation, we perform the left polar decompositions of the matrices $\tilde{E}$ and $\tilde{F}$:
\begin{equation}
      \tilde{E} = P_E V_{E}, ~ ~  \tilde{F} = P_F V_{F}.
\end{equation}
The matrices $V_{E}$ and $V_{F}$ are $N\times N(M+1)$ matrices with orthogonal rows.  
The matrices $P_E$ and $P_F$ are positive definite square $N\times N$ matrices. 
From the definition,
        $P_E =  (\tilde{E}\tilde{E}^H)^{\frac12}$ and $P_F =  (\tilde{F}\tilde{F}^H)^{\frac12}$.
According to Eq.~\eqref{commut_part}, the matrices $\tilde{E}\tilde{E}^H$ and $\tilde{F}\tilde{F}^H $ commute.
Therefore, the matrices $P_E$ and $P_F$ can be diagonalized simultaneously:
\begin{equation}
P_E  = U \Lambda_E U^H, ~ ~   P_F = U \Lambda_F U^H,
\end{equation}
where $\Lambda_E^2 = \Lambda_F^2 + \mathbf{1}_N$. 

As a result, the matrices $\tilde{E}$ and $\tilde{F}$ read
\begin{align}
        \tilde{E} = U \Lambda_E U^H V_{E} = U \Lambda_E W_E, \\
        \tilde{F} = U \Lambda_F U^H V_{F} = U \Lambda_F W_F,
\end{align}
that can be interpreted as a simultaneous singular value decomposition with the left unitary matrix~$U$.

In opposite to the full Bogoliubov transformation, the partial transformation describes an open system and is not invertible.
Thus, there is no `rotational condition', which exists for the full transformation, so the Bloch-Messiah reduction is not valid for open systems. 
Therefore, the basis~$U$ for partial Bogoliubov transformation is defined up to arbitrary phases for different modes.

\section{ Equations for continuous frequencies  }
\label{appendix_continuous_freq}
 
In this section, we briefly present equations for the continuous frequency space.
Taking the limit $T\rightarrow\infty$ in Eq.~\eqref{ap_eq_field_quantized}, one gets 
$T^{-1} \sum_{\omega_m} \rightarrow (2\pi)^{-1} \int d\omega$ 
and continuous operators are introduced as  $\hat{a}(\omega, z) \equiv \lim_{\Delta \omega \rightarrow0} \frac{ \hat{a}_i(z) }{\sqrt{\Delta \omega}}$.
Consequently, the Langevin equation has the form
 \begin{equation}
    \dfrac{d \hat{a}(z, \omega) }{d z} =  
    i \kappa(\omega) \hat{a}(z, \omega) \\ + i\Gamma \int d\omega^{\prime} J (z, \omega, \omega^\prime) \hat{a}^\dagger(z,\omega^\prime)  + \sqrt{\alpha(\omega)} \hat{f}(z,\omega),
    \label{eq_langevin_cont}
\end{equation}
where $J(z,\omega, \omega^\prime) = S(\omega+\omega^\prime)e^{i k_p(\omega+\omega^\prime) z}$ and $\kappa(\omega)=k(\omega)+i\alpha(\omega)/2$.
The field and environment operators now satisfy the continuous bosonic commutation relations $[\hat{a}(z,\omega),\hat{a}^\dagger(z,\omega^\prime)]=\delta(\omega-\omega^\prime)$ and  $[\hat{f}(z,\omega),\hat{f}^\dagger(z,\omega^\prime)]=\delta(\omega-\omega^\prime)$.

The solution to the Langevin equation reads
\begin{multline}
    \hat{a}(z, \omega) = \int d\omega^\prime \; \bigg[ g(z, z_0, \omega, \omega^\prime) \hat{a}(z_0, \omega^\prime)  +
     h(z, z_0, \omega, \omega^\prime) \hat{a}^\dagger(z_0, \omega^\prime)  \\ +
     \sqrt{\alpha(\omega) } \int_{z_0}^z \: dz^\prime \Big( g(z, z^\prime, \omega, \omega^\prime)\hat{f}(z^\prime, \omega^\prime)   +
     h(z, z^\prime, \omega, \omega^\prime) \hat{f}^\dagger(z^\prime, \omega^\prime) \Big)  \bigg], 
    \label{eq_langevin_cont_solution}
\end{multline}
where the functions $g(z, x, \omega, \omega^\prime)$ and $h(z, x, \omega, \omega^\prime)$ are determined by the equations 
\begin{equation}
    \partial_z g(z, x, \omega, \omega^\prime)  =  i\kappa(\omega) g(z, x, \omega, \omega^\prime)    + i\Gamma \int d\omega^{\prime\prime} \ J(z, \omega, \omega^{\prime\prime}) h^*(z, x, \omega^{\prime\prime}, \omega^\prime),
    \label{eq_func_g_cont}
\end{equation}
\begin{equation}
    \partial_z h(z, x, \omega, \omega^\prime)  =  i\kappa(\omega) h(z, x, \omega, \omega^\prime)    + i\Gamma \int d\omega^{\prime\prime} \ J(z, \omega, \omega^{\prime\prime}) g^*(z, x, \omega^{\prime\prime}, \omega^\prime),
    \label{eq_func_h_cont}
\end{equation}
with the initial conditions $g(x, x, \omega, \omega^\prime)=\delta(\omega-\omega^\prime)$ and $h(x, x, \omega, \omega^\prime)=0$.
For $\alpha(\omega)\equiv 0$, Eqs.~\eqref{eq_func_g_cont} and \eqref{eq_func_h_cont} correspond to the well-known solution for the lossless medium (see e.g. Refs.~\cite{Christ_2013,Sharapova_2020}).

For the Gaussian initial state and the Markovian Gaussian environment with 
 $\braket{\hat{f}^\dagger(z, \omega)} = \braket{\hat{f}(z, \omega)} = 0$ and the correlation matrices $\braket{\hat{f}^\dagger(z,  \omega) \hat{f}(z^\prime,  \omega^\prime)} = \braket{\hat{f}^\dagger(z,  \omega) \hat{f}(z,  \omega^\prime)}\delta(z-z^\prime)$ and $\braket{\hat{f}(z,  \omega) \hat{f}(z^\prime,  \omega^\prime)} = \braket{\hat{f}(z,  \omega) \hat{f}(z,  \omega^\prime)}\delta(z-z^\prime)$, the corresponding master equations have the form  
 \begin{multline}
\dfrac{d \braket{\hat{a}^\dagger(z, \omega) \hat{a}(z, \omega^\prime)}}{d z} = 
        \big( -i \kappa^*(\omega) + i \kappa(\omega^\prime)    \braket{\hat{a}^\dagger(z, \omega) \hat{a}(z, \omega^\prime)}  + 
           i\Gamma \int d\omega^{\prime\prime} J (z, \omega^\prime, \omega^{\prime\prime})   \braket{\hat{a}^\dagger (z, \omega) \hat{a} (z, \omega^{\prime\prime})} \\
         - i\Gamma \int d\omega^{\prime\prime} J^* (z, \omega, \omega^{\prime\prime}) \braket{\hat{a}(z, \omega^{\prime\prime}) \hat{a}(z, \omega^\prime)} 
        + \sqrt{\alpha(\omega)  \alpha(\omega^\prime)}  \braket{\hat{f}^\dagger(z,\omega) \hat{f}(z,\omega^\prime)},
        \label{eq_correlation_equation_1_cont}
\end{multline}
\begin{multline}
    \dfrac{d \braket{\hat{a}(z, \omega) \hat{a}(z, \omega^\prime)}}{d z} = \big( i \kappa(\omega) + i \kappa(\omega^\prime) \big)   \braket{\hat{a}(z,\omega)\hat{a}(z, \omega^\prime )} 
    + i\Gamma  \int d\omega^{\prime\prime} J(z, \omega^\prime, \omega^{\prime\prime}) \braket{\hat{a}(z, \omega)\hat{a}^\dagger(z,\omega^{\prime\prime})} \\ 
    + i\Gamma  \int d\omega^{\prime\prime} J(z, \omega, \omega^{\prime\prime}) \braket{\hat{a}^\dagger(z,\omega^{\prime\prime})\hat{a}(z, \omega^\prime) } 
    + \sqrt{\alpha(\omega)  \alpha( \omega^\prime)} \braket{\hat{f}(z, \omega) \hat{f}(z, \omega^\prime)}. 
    \label{eq_correlation_equation_2_cont}
\end{multline}

\section{ Refractive indices of the model  }
\label{appendix_ref_index}

The refractive index for the PDC field ($o$-wave) is taken as
\begin{equation}
  n_o^2(\lambda)=2.7359 + \dfrac{0.01878}{\lambda^2 - 0.01822} - 0.01354 \lambda^2.
\end{equation}
The refractive index for the pump field ($e$-wave) is
\begin{equation}
  n_e^2(\lambda)= \Bigg( \dfrac{\sin^2(\theta)}{ \eta^2(\lambda) } + \dfrac{\cos^2(\theta)}{n_o^2(\lambda)}    \Bigg)^{-1},
\end{equation}
where
\begin{equation}
  \eta^2(\lambda)=2.3753 + \dfrac{0.01224}{\lambda^2 - 0.01667} - 0.01516 \lambda^2,
\end{equation}

 $\theta=0.1107\pi$ and $\lambda$ should be taken in $\mu$m.

\end{document}